\documentclass[fleqn,usenatbib]{mnras}

\usepackage{newtxtext,newtxmath}
\usepackage{tikz}
\usepackage{pgfplots}
\usetikzlibrary{arrows.meta, positioning, calc, angles, quotes, math, intersections, fillbetween, decorations.markings, shapes.geometric}
\DeclareMathOperator{\erf}{\rm{erf}} 

\usepackage[T1]{fontenc}

\DeclareRobustCommand{\VAN}[3]{#2}
\let\VANthebibliography\thebibliography
\def\thebibliography{\DeclareRobustCommand{\VAN}[3]{##3}\VANthebibliography}

\usepackage{graphicx}	
\usepackage{amsmath}	
\usepackage{bm}
\usepackage{multirow}
\usepackage{comment}
\usepackage[table]{xcolor}

\usepackage{scalerel,tikz}
\usetikzlibrary{svg.path}
\definecolor{orcidlogocol}{HTML}{A6CE39}
\tikzset{orcidlogo/.pic={
 \fill[orcidlogocol] svg{M256,128c0,70.7-57.3,128-128,128C57.3,256,0,198.7,0,128C0,57.3,57.3,0,128,0C198.7,0,256,57.3,256,128z};
 \fill[white] svg{M86.3,186.2H70.9V79.1h15.4v48.4V186.2z}
 svg{M108.9,79.1h41.6c39.6,0,57,28.3,57,53.6c0,27.5-21.5,53.6-56.8,53.6h-41.8V79.1z M124.3,172.4h24.5c34.9,0,42.9-26.5,42.9-39.7c0-21.5-13.7-39.7-43.7-39.7h-23.7V172.4z}
 svg{M88.7,56.8c0,5.5-4.5,10.1-10.1,10.1c-5.6,0-10.1-4.6-10.1-10.1c0-5.6,4.5-10.1,10.1-10.1C84.2,46.7,88.7,51.3,88.7,56.8z};
}}
\newcommand\orcidicon[1]{\href{https://orcid.org/#1}{\mbox{\scalerel*{
\begin{tikzpicture}[yscale=-1,transform shape]
\pic{orcidlogo};
\end{tikzpicture}
}{|}}}}

\title[Erasure of bar resonances by dark subhaloes]{The erasure of Galactic bar resonances by dark matter subhaloes}

\author[E. Y. Davies et al.]{Elliot Y. Davies~\orcidicon{0000-0001-5996-4072}$^{1}$\thanks{E-mail: eydavies@mit.edu}, Adam M. Dillamore~\orcidicon{0000-0003-0807-5261}$^{2}$, Vasily Belokurov~\orcidicon{0000-0002-0038-9584}$^{3}$ and Lina Necib~\orcidicon{0000-0003-2806-1414}$^{1}$ .
\\
$^{1}$MIT Kavli Institute For Astrophysics and Space Research, 70 Vassar St, Cambridge, MA 02139, USA \\
$^{2}$Department of Physics and Astronomy, University College London, London, WC1E 6BT, UK \\
$^{3}$Institute of Astronomy, University of Cambridge, Madingley Road, Cambridge CB3 0HA, UK
}
\date{Accepted XXX. Received YYY; in original form ZZZ}

\pubyear{\the\year{}}

\begin{document}
\label{firstpage}
\pagerange{\pageref{firstpage}--\pageref{lastpage}}
\maketitle

\begin{abstract}
In the context of increasing appreciation for the coupling between the Galactic bar and halo, we introduce a new framework using stars trapped in resonance with the bar to probe the Galactic dark matter subhalo population. Since resonant stars occupy a finite width in action space, perturbations from subhaloes can shift a star's actions beyond this width, causing them to circulate out of resonance. Physically, the dark substructure in the Milky Way may dissolve, puff-up, or re-order the resonant features in the stellar halo. To explore the utility of this framework, we treat individual encounters in the impulse approximation and model their cumulative effect as diffusion in the relevant action. The resulting diffusion coefficient allows us to link the survival of resonant populations to the subhalo mass function, whose properties depend on the particle nature of dark matter. Test-particle integration validates the impulse treatment for low-mass subhaloes and quantifies its regime of applicability. For a Milky Way-like bar, we find individual subhaloes with $M < 10^7\,{\rm M_\odot}$ have negligible impact on stars in co-rotation resonance, whereas the full cold dark matter (CDM) population could erase the resonance over the bar’s lifetime. The persistence of resonances therefore implies a suppression of the local subhalo density to less than $1/3$ of CDM expectations, consistent with tidal disruption and previous literature. The narrower widths of higher-order resonances will increase the constraining power of this framework, and therefore motivates searches for bar-resonant features in observational data.
\end{abstract}

\begin{keywords}
Galaxy: kinematics and dynamics -- Galaxy: centre -- (cosmology:) dark matter
\end{keywords}



\section{Introduction}

In the era of precision astrometric instruments such as \textit{Gaia} \citep[][]{Gaia}, the Milky Way provides a uniquely detailed laboratory for testing the nature of dark matter (DM). The most widely accepted model of DM -- the so-called \textit{cold} dark matter (CDM) model -- predicts a hierarchical structure of \textit{subhaloes} contained within the larger dark matter halo of any Milky Way sized galaxy \citep[e.g.][]{blumenthal1984formation, white1991galaxy}. In the CDM framework, these haloes-within-haloes are theorized to have masses ranging from larger than a dwarf galaxy to smaller than the Earth. While the presence of dark matter haloes in Galactic dwarf galaxies (such as the Large Magellanic Cloud, Fornax and Sculptor) is now incontrovertible, there has yet to be confident detection of any subhalo on the lower end of the mass-spectrum. Detection of subhaloes below $M \lesssim 10^8$ M$_{\odot}$ are crucial for discerning whether the actual population of Galactic subhaloes deviates from CDM expectations; alternative dark matter models predict a suppression at the lower end of the mass-spectrum \citep[e.g.][]{schneider2013halo, lovell2014properties, kulkarni2022what}. However, it is at these lower masses where subhaloes struggle to maintain a bound stellar population, and thus chart invisible trajectories through the Galaxy \citep[][]{read2017stellar, jethwa2018upper, nadler2020milky, kravtsov2022grumpy}. Therefore, the challenge to the community is to measure perturbations to the stellar material in the Galaxy from these low-mass, non-luminous subhaloes as a means of indirect detection.
 
While the aforementioned CDM model may be the most prolific model in astronomy, its small-scale shortcomings \citep[e.g. see][for review]{bullock2017small} and the lingering lack of direct detection are good cause to explore alternatives, even with baryonic nuances considered \citep[e.g.][]{governato2010bulgeless, madau2014dark}. Often discussed shortcomings include the discrepancy between observed dark matter dominated cored haloes \citep[e.g.][]{flores1994observational, moore1994evidence, deblok2001high, walker2011method, oh2015high-res} and simulated dark matter only cuspy haloes \citep[e.g.][]{navarro1996structure, navarro1997universal, moore1999dark, diemand2005cusps} (dubbed the \textit{cusp-core problem}), and the unexpected diversity of dwarf galaxy rotation curves \citep[][]{oman2015unexpected, kaplinghat2019too}. Scrutinising the low end of the subhalo mass function therefore remains a worthwhile probe of alternative models. In this work we discuss two out of the numerous alternatives to CDM: warm dark matter (WDM) and self-interacting dark matter (SIDM). 

WDM is a general class of dark matter models in which the particle is deemed lighter than CDM \citep[e.g.][]{bond1982formation, bond1983collionless, blumenthal1984formation, bode2001halo}, with sterile neutrinos introduced as an early (but now well-constrained) candidate \citep[e.g.][]{dodelson1994sterile}. Specifically, with the lower bound on CDM taken to be roughly $m_{\rm CDM} > \mathcal{O}(10~\rm keV)$, current astrophysical bounds permit the WDM to have a mass of approximately $m_{\rm WDM}~\sim~3-5$~keV \citep[e.g.][]{irsic2017new}. This difference in mass simply grants the particle a longer free streaming distance, thereby washing out structure on small scales. This translates to a reduction in power at the low end of the subhalo mass function. Refined measurements of the subhalo mass function would, therefore, enable the community to even more precisely narrow the window of acceptable particle masses in the WDM-CDM realm. 

Interest in SIDM, a model in which the dark matter particles can scatter with each other, has continued since its inception by \citet[][]{carlson1992self}. The scattering allows heat transport within a halo which can transform cuspy profiles into cored profiles \citep[e.g.][]{spergel2000observational}, and eventually into \textit{core collapsed} objects \citep[see e.g.][for a review]{tulin2018dark}. This kind of heat transport is not permitted in the traditional collisionless models of CDM and WDM. The properties of SIDM are well-motivated by ``dark sector'' extensions the standard model of particle physics \citep[e.g.][]{ackerman2009dark, arkani-hamed2009theory}. Proponents of SIDM argue that it alleviates many of the small scale issues of CDM, while matching the large scale successes \citep[e.g.][]{kaplinghat2016dark}. Recent work finds that lensing of the inner density of galaxy clusters reveal properties inconsistent with CDM \citep[][]{natarajan2026new}. They argue core-collapse resulting from SIDM better explains the discovered steep density slopes. Complementary to previous work, strong gravitational lensing has revealed a dark structure which models suggest has a density profile which favours SIDM, with CDM predicting a value an order of magnitude too small \citep[][]{vegetti2026possible}. 

The utility of the Milky Way as a probe of different dark matter models (or other cosmological phenomena) is commonly referred to as ``near-field cosmology''. One of the now canonical examples of a resource in the near-field cosmology toolbox is the use of cold stellar streams \citep[][]{ibata2002uncovering, johnston2002how}. These elongated stellar features are the last stage of evolution of a globular cluster (GC) or dwarf galaxy undergoing tidal disruption. Their wide spatial extent yet narrow velocity dispersion ($\sigma \lesssim 1$ km/s, in the case of a GC stream) means that impacts by low mass subhaloes leave a persistent mark on their spatial distribution, colloquially called ``gaps'' \citep[][]{carlberg2009star, erkal2015forensics, erkal2015properties}. The evolution of a stream on a well-modelled orbit, after impact from a subhalo, is well understood. Moreover, in the case of the stream GD-1 \citep[][]{grillmair2006detection, pricewhelan2018off}, the impact from a low-mass subhalo may have already been detected \citep[][]{bonaca2019spur, nibauer2025measurement}. The difficulty that the stream-subhalo method now faces is the inherent messy, non-equilibrium state of the Milky Way's potential. The non-axisymmetry of the global potential \citep[][]{han2022stellar, han2022tilt, dillamore2025geometry}, as well as non-dark interactions from Giant Molecular Clouds \citep[e.g.][]{amorisco2016gaps, pearson2017gaps} are very difficult to disentangle from possible impacts with low mass subhaloes. To get a bound on the low-mass Galactic subhalo population, it may be necessary to take a wider approach \citep[e.g.][]{davies2023ironing, forouhar2025accreted}, rather than looking for interactions on a case-by-case basis. To complement searches for the imprint of individual subhaloes on streams, it is important (yet theoretically and numerically difficult) to explore the global heating of stellar debris from the entire population of low mass subhaloes. This framework clearly comes with its own challenges. The difficulty in numerical modelling is made clear by \citet[][]{ludlow2023spurious}. They illustrate the difficulty in disentangling numerical artifacts (caused by mass resolution) from real heating. Moreover, \citet[][]{penarrubia2019stochastic} discusses the importance of small perturbers in similar contexts, noting the challenges of modelling the evolution of any gravitating system embedded in a lumpy medium.

Stellar streams are often viewed as isolated subhalo detectors; the current zeitgeist in Galactic studies is to examine one stream at a time for the influence of a single (or a few) subhaloes \citep[e.g.][]{bonaca2019spur}. However, over a hundred stellar streams have now been found scattered across the Milky Way's halo, forming a tapestry of Galactic dark matter detectors \citep[][]{belokurov2006field, malhan2018streamfinderI, mateu2023galstreams, bonaca2024stellar, shih2024viamachinae2}. This ensemble of streams could plausibly allow a characterisation of what may be termed the Galactic subhalo background, by some measure of baseline heating (i.e. velocity dispersion) across the entire stream population. While the simulation of large numbers of low mass subhaloes is costly, tools such as StreamSculptor \citep[][]{nibauer2025streamsculptor} make this kind of study more tractable.

This work argues that a currently under-utilised resource in near-field cosmology toolbox is the Galactic bar. Despite the difficulty of mapping the inner Galaxy, it is extremely likely that the Milky Way has a bar. Evidence for a Galactic bar comes from gas kinematics \citep[][]{peters1975models, binney1991understanding}, near-infrared emission \citep[][]{blitz1991direct}, and stellar kinematics \citep[][]{howard2009kinematics, shen2010milky, debattista2017separation}. While the properties of the bar are yet to be pinned-down exactly, its radial length has been measured to be around $3.5$ kpc \citep[e.g.][]{lucey2022constraining} and its pattern speed $\Omega_{\rm p}$ is estimated to be around $35 - 40$ km s$^{-1}$ kpc$^{-1}$ \citep[][and references therein]{shen2020bar}. A critical consequence of a non-axisymmetric contribution to the potential is the presence of resonances, i.e. regions where stellar orbital frequencies are commensurate with the bar’s rotation \citep[e.g.][]{lynden-bell1972generating, lynden-bell1979mechanism, kalnajs1991pattern, sellwood2010recent}. These resonances trap stars, reshape their orbits, and imprint coherent structures in phase space. Resonances such as the co-rotation, inner, and outer Lindblad resonances are thought to shape a number of observed structures in the Milky Way disk. This includes the Hercules moving group \citep[][]{dehnen1998distribution, dehnen2000effect} and other kinematic ridges found more recently in \textit{Gaia} data \citep[e.g.][]{antoja2018dynamically, trick2019galactic}.

While it has been long established that the bar at the centre of the Milky Way imparts resonant features in the disk, it has been shown that this effect may be responsible for observed kinematic overdensities in the Galactic halo \citep[][]{moreno2015resonant, dillamore2023stellar, dillamore2024radial}. Consequently, bar-induced resonant features are likely to interact with the dark matter halo. Understanding this link between the Milky Way’s dark halo and the Galactic bar is crucial, as angular-momentum exchange between these components impacts both the evolution of the bar and the response of the inner halo. For example, it has been shown that the dynamical friction from the halo causes the bar to slow down \citep[e.g.][]{chiba2022oscillating}. However, the consideration of gas complicates this picture and may prevent any significant decrease in the pattern speed \citep[][]{beane2023stellar}.

Amongst the growing literature on the link between the dark content of the Galaxy and the bar, we propose a new method for probing the Galactic subhalo background with the resonant features induced by the bar itself. Simply put, if the bar creates resonant features in the halo, the Galactic subhalo background may leave an imprint on these features. Such an imprint would be either the complete dissolution of an expected resonant population, enhanced velocity dispersion of a resonance, or perhaps an overpopulation of higher order resonances due to heating of lower order resonances. This work seeks to be a first exploration of the viability of this method, by means of an analytical model and test particle simulations. 

The outline of this work is as follows. In Sec.~\ref{sec:mathe_background} we detail the mathematical background behind bar resonances. In Sec.~\ref{sec:single_subhalo} we explore the impact of a single subhalo using impulse approximation, contextualised with test particle simulation. In Sec.~\ref{sec:many_subhaloes} we show the impact of a population of subhaloes, modelled by a diffusive random walk. In Sec.~\ref{sec:discussion}, we present and discuss the results. Lastly, we summarise this paper and consider future work in Sec.~\ref{sec:summary}

\section{Bar resonances}\label{sec:mathe_background}

In this section, we review the necessary mathematical background to understand the resonant behaviour resulting from the bar, and find the maximum change in action $\Delta I_{\rm half}$ required for a resonant star to be ejected out of resonance along its orbit. In this first paper, we focus entirely on circular orbits and leave halo-like orbits for future work. For other resources that cover the relevant resonant dynamics thoroughly, see the works of \citet[][]{chiba2021resonance}, \citet[][]{chiba2022oscillating} and \citet[][]{hamilton2023galactic}.

\subsection{Galactic model}

Throughout this work we take the potential to comprise two components. Namely, we describe the Milky Way using a cored logarithmic potential with core radius $r_{\rm core}$ and peak velocity $v_0$,
\begin{equation}\label{logarithmic}
    \Phi_0(r) = \frac{v_0^2}{2}\log(r_{\rm core}^2 + r^2),
\end{equation}
which is then perturbed by a bar potential, with pattern speed $\Omega_{\rm p}$, of the form
\begin{equation}\label{eq:bar_potential}
    \delta \Phi(r, \varphi, t) = \Phi_{\rm b}(r)\cos[2(\varphi - \Omega_{\rm p}t)],
\end{equation}
where the radial component is
\begin{equation}\label{bar_radial}
    \Phi_{\rm b}(r) = -\frac{Av_0^2}{2}\frac{r^2}{(r + r_{\rm b})^5}.
\end{equation}
Here $v_0$ is the velocity of the rotation curve (in the limit of $r\gg r_{\rm core}$) in the logarithmic potential in Eq.~\ref{logarithmic}, $r_{\rm b}$ is the scale length of the of the bar (not equivalent to the actual length), and the coefficient $A$ is the ratio of the azimuthal force due to the bar and the radial force due to the unperturbed potential at the radius of co-rotation $r=r_{\rm CR}$
\begin{equation}
    A \equiv \frac{\left|\frac{1}{r}\frac{d\Phi_{\rm b}}{d\varphi}\right|_{r_{\rm CR}}}{\left| \frac{d\Phi_0}{dr}\right|_{r_{\rm CR}}},
\end{equation}
where the co-rotation radius is the distance from the Galactic centre where a star rotates at the same angular speed of the bar i.e. where $v_{\rm circ}(r) = r \, \Omega_{\rm p}$. Altogether, this implies a Hamiltonian of the form
\begin{equation}\label{hamiltonian}
    H = \frac{1}{2}|\bm{v}|^2 + \Phi_0 + \delta \Phi = H_0 + \delta \Phi,
\end{equation}

\subsection{Action-angle variables}

For any integrable Hamiltonian $H(\bm{q}, \bm{p})$ that is a function of generalised coordinates $(\bm{q}, \bm{p})$, we can perform a canonical transformation to rewrite the Hamiltonian in terms of a set of useful integrals of motions called \textit{actions} $\bm{J}$ and their conjugate coordinates known as \textit{angles} $\bm{\theta}$ which evolve linearly in time. Hamilton's equations in these coordinates are
\begin{equation}
    \dot{\bm{J}} = - \frac{\partial H}{\partial \bm{\theta}} =0 , \:\:\: \dot{\bm{\theta}} = \frac{\partial H}{\partial \bm{J}} = \bm{\Omega}(\bm{J}).
\end{equation}
Where $\bm{\Omega} = \dot{\bm{\theta}}$ are the frequencies associated with the coordinates. The actions are defined by integrating the generalised coordinates via
\begin{equation}\label{definition_action}
    J_i = \frac{1}{2\pi}\oint p_i dq_i.
\end{equation}
These coordinates are especially useful in the context of studying the orbits in the Milky Way with a rotating central bar, as in the Hamiltonian in Eq.~{\ref{hamiltonian}}. 

\subsection{Resonance Hamiltonian}

A \textit{resonance} with the Galactic bar occurs when an integer combination of these orbital frequencies $\bm{\Omega}$ is commensurate with the rotation frequency (or ``pattern speed'') of the bar $\Omega_{\rm p}$:
\begin{equation}\label{res_condition_omega}
    \bm{n}\cdot\bm{\Omega} = n_{\varphi}\Omega_{\rm p},
\end{equation}
where $\bm{n} =(n_{r}, n_{\varphi}, n_z) \in \mathbb{Z}^3$. The integers $n_r, n_{\varphi}, n_z$ specify the order of the resonance in each degree of freedom. The unperturbed Hamiltonian $H_0$ describing the Milky Way (where the bar is treated as a perturbation) permits action-angle coordinates in the radial direction, azimuthal and vertical directions. We label the angles as $\bm{\theta} = (\theta_r, \theta_{\varphi}, \theta_z)$ and the actions as $\bm{J}=(J_r, L_z, J_z)$. The corresponding frequencies are thus $\Omega(\bm{J}) = (\Omega_r, \Omega_{\varphi}, \Omega_z)$. However, in the context of studying resonant orbits of the bar, it is convenient to define the \textit{slow frequency}, 
\begin{equation}
    \Omega_{\rm slow} \equiv \bm{n}\cdot\bm{\Omega} - n_{\varphi}\Omega_{\rm p}
\end{equation}
which is evidently $\Omega_{\rm slow} \approx 0$ near resonance. The time integral of the slow frequency is the slow angle (i.e. the phase of star relative to the bar):
\begin{align}
    \phi &= \bm{n}\cdot\bm{\theta} - n_{\varphi}\Omega_{\rm p}t \nonumber \\ 
    &= n_r\theta_r + n_{\varphi}\theta_{\varphi} - n_{\varphi}\Omega_{\rm p}t,
\end{align}
where we have set $\bm{n} = (n_r,n_{\varphi},0)$ because the logarithmic potential and bar component are primarily acting in the radial and azimuthal directions, hence the vertical motion will not strongly couple to the bar.  By appropriate choice of some \textit{fast angle} $\tilde{\phi}$, we can make a canonical transformation from the spherical coordinate action-angles $(\bm{\theta}, \bm{J})$ to some ``fast and slow'' action-angles $(\bm{\theta}', \bm{J}')$, where $\bm{\theta}' = (\phi, \tilde{\phi})$ are the slow and fast angles and $\bm{J}'=(I,\tilde{I})$ are the slow and fast actions. We define the fast angle to be equal to the radial angle $\tilde{\phi} =\theta_r$. We can find these new canonical actions $\bm{J}'$ from
\begin{equation}
    \bm{J} = \frac{\partial S}{\partial \bm{\theta}}, \:\:\: \bm{\theta}' = \frac{\partial S}{\partial \bm{J}'}
\end{equation}
for a canonical transformation with a generating function $S$, where the new Hamiltonian is given by
\begin{equation}
    \bar{H}(\bm{\theta}', \bm{J}', t) = H(\bm{\theta}, \bm{J}, t) + \frac{\partial S}{\partial t},
\end{equation}
and the simplest generating function $S$ is
\begin{equation}
    S(\bm{\theta} , \bm{J}', t) = \left[n_r\theta_r + n_{\varphi}\theta_{\varphi} - n_{\varphi} \Omega_{\rm p} t\right]I + \theta_r\tilde{I}.
\end{equation}
The conjugate action to this slow angle is the \textit{slow action} (i.e. how far the star’s frequencies deviate from perfect commensurability with the bar's pattern speed) can be found using the generating function:
\begin{equation}
    J_{\varphi} = \frac{\partial S}{\partial \theta_{\varphi}} = n_{\varphi}I.
\end{equation}
Then finally, re-labelling $L_z = J_{\varphi}$, the slow action for the co-rotation resonance is given by,
\begin{align}
    I = L_z / n_{\varphi}.
\end{align}
Using the equation for the new Hamiltonian, and expanding the perturbation in Fourier series with coefficients $\Psi_k (\bm{J}', t)$, we can show that,
\begin{equation}
    \bar{H}(\bm{\theta}', \bm{J}', t) = H_0(\bm{J}') + \sum_{\bm k} \Psi_{\bm k} (\bm{J}', t)\exp{(i\bm{k}\cdot\bm{\theta}')} - m\Omega_{\rm p}I.
\end{equation}
As is done by \citet[][]{hamilton2023galactic}, we subsequently average over the fast motion, and Taylor expand for small $I$ near the resonance $I_{\rm res}$. Averaging over the fast angle results in a new Hamiltonian:
\begin{equation}
    \mathcal{H} = H_0(I) + 2\sum_{k} \Psi_{(0,k)}(I) \exp (ik\phi) - m\Omega_{\rm p}I.
\end{equation}
It is convenient to define a shifted version of the slow angle $\upphi$ such that $\upphi \in (-\pi,\pi)$:
\begin{equation}
    \upphi = \phi + \frac{\arg \Psi_{(0,k)}}{k}.
\end{equation}
However, as shown by \citet[][]{chiba2022oscillating}, only the $k=1$ term contributes for resonances where $n_{\varphi} = 2$. 
This ultimately results in the pendulum Hamiltonian  $\mathcal{H}$ for resonant motion (ignoring higher order terms and constants):
\begin{equation}\label{eq:resonance_hamiltonian}
    \mathcal{H} = \frac{1}{2}\alpha(I-I_{\rm res})^2 - \epsilon\cos(\upphi)
\end{equation}
The resulting resonance Hamiltonian is evidently only a function of the slow angle $\phi$ and the slow action $I$. The coefficient $\alpha$ is given by
\begin{equation}
    \alpha = \frac{\partial^2 H_0}{\partial I^2} \bigg\rvert_{I_{\rm res}},
\end{equation}
and the coefficient $\epsilon$ is given by
\begin{equation}\label{eq:epsilon}
    \epsilon = 2|\Psi_{(0,1)}(I)| = |\Phi_{\rm b}(r_{\rm res})|.
\end{equation}
The second equality is valid for the co-rotation resonance when $J_r \rightarrow 0$, as shown by \citet[][Appendix B]{chiba2021resonance}. The contours of Eq.~\ref{eq:resonance_hamiltonian} can be seen in Fig.~\ref{fig:separatrix}. The dashed black lines show the separatrix -- the transition point from one behaviour to another -- between librating orbits and circulating orbits. The black point in the centre shows the exact resonance $(I,\phi) = (I_{\rm res},0)$, and the red arrow shows the half width in slow action $\Delta I_{\rm half}$. The black solid lines show examples of librating and circulating orbits.

We can use Hamilton's equations to rewrite $\alpha$ in a more practical form:
\begin{equation}
    \alpha = \frac{\partial\bm{(\bm{n}\cdot\Omega})}{\partial I}\bigg\rvert_{I_{\rm res}}.
\end{equation}
In a logarithmic potential (for circular orbits at $r \gg r_{\rm core}$), where $\Omega_r = \sqrt{2}v_0^2/L_z$ and $\Omega_{\varphi}=v_0^2/L_z$ are functions of $L_z$ only, we can write $\alpha$ as (keeping the fast action $\tilde{I}$ constant),
\begin{equation}\label{alpha}
    \alpha = \frac{\partial}{\partial(L_z/n_{\varphi})}\left(n_{\varphi}\Omega_{\varphi} + n_r\Omega_r\right) = -n_{\varphi}(n_{\varphi}+n_r\sqrt{2})\frac{v_0^2}{L_z^2}.
\end{equation}
This expression is only exact for the co-rotation resonance $(n_r=0)$ in our toy model. However, we use it as a rough estimate of the relative widths of nearby resonances (see Table~\ref{tab:resonances}). A fully consistent treatment for $n_r \neq 0$, in which the slow action mixes $J_r$ and $J_{\varphi}$ is given in \citet[][Appendix C]{chiba2022oscillating}. For $r\gg r_{\rm core}$ the radius of the resonance is related to the bar pattern speed and flat rotation curve velocity $v_0$ by,
\begin{equation}\label{eq:res_radius}
    r_{\rm res}(n_r,n_{\varphi}) = \frac{v_0}{\Omega_{\rm p}}\left(1+\frac{n_r}{n_{\varphi}}\sqrt{2}\right),  
\end{equation}
which is found from Eq.~\ref{res_condition_omega}, and the definition for $\Omega_{\varphi}(R) = v_0 / R$ in a logarithmic potential for circular orbits.

\subsection{Resonance separatrix}

\begin{figure}
    \centering
    \includegraphics[width=0.95\columnwidth]{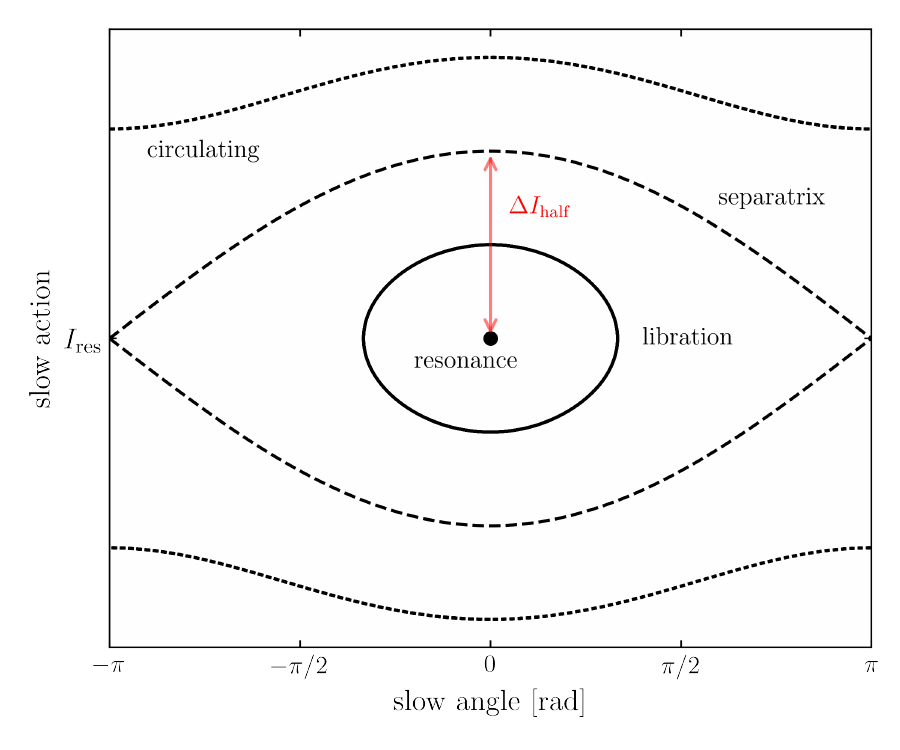}
    \caption{Sketch of the contours of the bar resonance Hamiltonian (Eq.~\ref{eq:resonance_hamiltonian}) shown in the slow angle $\upphi$ and slow action $I$ space. The central point is the location of the parent orbit, exactly in resonance. The inner solid black contour is an example libration orbit around the resonance. The black dashed line is the separatrix between resonant and circulating orbits. The outer dotted black contour is an example circulating orbit. The red arrow indicates the change in slow action $I$ required to kick the parent orbit beyond the separatrix.}
    \label{fig:separatrix}
\end{figure}

The survival of bar-induced resonance features depends on whether or not stars remain confined to a finite region of action space (see Fig.~\ref{fig:separatrix}). In this work, we seek to learn whether dark matter subhaloes (either individually or as a population) are capable of moving stars out of this region, causing them to lose coherence with the bar. Therefore, we must characterise the boundary in action space that separates resonantly trapped orbits from non-trapped orbits. The boundary is defined by the \textit{separatrix} of the resonance Hamiltonian. Stars whose slow actions lie within the separatrix remain trapped, while stars pushed beyond the separatrix circulate freely. Here, we detail the structure of the resonance orbit families and derive the width of the separatrix i.e. the minimum perturbation required for a star to escape resonance.. 

A ``family'' of resonant orbits trapped at a given $n_r\!:\!n_{\varphi}$ resonance consists of one closed, periodic \emph{parent orbit} and a surrounding collection of \emph{librating orbits}. The parent orbit closes in the frame co-rotating with the bar’s pattern speed after completing $n_{\varphi}$ azimuthal revolutions and $n_r$ radial oscillations, forming the backbone of the resonance. In this rotating frame, the parent orbit maintains a constant slow angle $\upphi$ and lies close to the $n_r\!:\!n_{\varphi}$ resonance line in action space. For example, the aforementioned co-rotation corresponds to $(n_{\varphi},n_r)=(2,0)$, whereas the outer Lindblad resonance (OLR) corresponds to $(n_{\varphi},n_r)=(2,1)$. Higher order resonances appear at larger radii and are progressively weaker, but they can still influence stellar phase-space structure.

The librating orbits represent stars that are nearly, but not exactly, on this parent trajectory. Their motion remains phase locked to the bar, so that their slow angle $\upphi$ oscillates gently around the parent orbit’s value. In action space, this corresponds to oscillations about the resonant slow action $I_{\rm res}$. Each librating orbit differs mainly by the amplitude of this slow oscillation, whose range of values increases with the strength of the bar. The largest such orbit marks the edge of the resonant region beyond which stars are no longer trapped and their phases drift freely with respect to the bar (i.e. the separatrix).

This structure can be visualized in Fig.~\ref{fig:separatrix}, where contours of constant $\mathcal{H}$ are plotted in the space of slow angle~$\upphi$ and slow action~$I$. In this diagram, the solid black line closed loops around the stable equilibrium point correspond to the librating orbits that remain trapped in resonance, while the dashed black line open contours represent circulating orbits whose phases drift relative to the bar. The black dashed line marks the separatrix. Stars inside the separatrix move coherently with the bar’s pattern whereas for those outside, the influence of the bar is averaged out over their orbit. 

The equations of motion for the slow angle and slow action can be obtained by Hamilton's equations:
\begin{equation}
    \dot{\upphi} = \frac{\partial \mathcal{H}}{\partial I} = \alpha(I-I_{\rm res}) \Rightarrow\ddot{\upphi}=\alpha\dot{I},
\end{equation}
where $\dot{I}$ is given by the other equation,
\begin{equation}
    \dot{I}=-\frac{\partial \mathcal{H}}{\partial \upphi}=-\epsilon\sin\upphi,
\end{equation}
The resulting equation of motion for the slow angle in the regime of small oscillations around resonance is
\begin{equation}
    \ddot{\upphi} + \alpha \epsilon\upphi = 0,
\end{equation}
taking $\upphi$ as small to utilise $\sin \upphi \sim \upphi$. As a reminder, $\upphi$ is actually shifted from the slow angle by $\pi$. From this equation of motion we can read off the libration frequency $\omega_{\rm lib}$ and libration timescale $\tau_{\rm lib}$ as
\begin{equation}
    \tau_{\rm lib} = \frac{2\pi}{\omega_{\rm lib}} = \frac{2\pi}{\sqrt{|\alpha\epsilon}|}. 
\end{equation}
The equilibrium points of the resonance Hamiltonian $\mathcal{H}$ are at the points where the derivates of the resonance hamiltonian with respect to both coordinates $I$ and $\upphi$ are zero, which are at $(I,\phi) = (I_{\rm res}, 0)$ and $(I,\phi) = (I_{\rm res}, \pi)$. Using these equilibrium points, we can find the width at which the separatrix ``half-width'' is greatest, i.e. the maximum of $|I-I_{\rm res}|$ over $\upphi$. Evaluating the resonance Hamiltonian at $\mathcal{H} = \pm \epsilon$, and rearranging for the slow action results in an equation describing the separatrix boundary:
\begin{equation}
    I(\upphi) = I_{\rm res} \pm \sqrt{\frac{4\epsilon}{|\alpha|}}|\cos(\upphi/2)|
\end{equation}
From this we get the half width:
\begin{equation}
    \Delta I_{\rm half} \equiv \max_{\phi} |I-I_{\rm res}| = \sqrt{\frac{4\epsilon}{|\alpha|}}.
\end{equation}
This half-width defines the maximum change in slow action required for a subhalo encounter to eject a star from resonance anywhere in its orbit. Therefore, if the parent orbit is moved in action space by an amount $\Delta I_{\rm half}$, it will be guaranteed to be removed from resonance. We can use Eq.~\ref{eq:epsilon} and Eq.~\ref{alpha} to evaluate $\Delta I_{\rm half}$ for any resonance and any bar properties. This expression neglects any additional restoring effects that would modify the separatrix width by a multiplicative factor. For a single impulsive encounter, however, such restoring forces operate on much longer timescales than the interaction itself and therefore do not affect the instantaneous ejection criterion. This is likely not true for many subhaloes. In Table.~\ref{tab:resonances} we show the value of a range of half-widths for a few different resonances for comparison with the co-rotation resonance. As well as the aforementioned co-rotation and Outer Lindblad resonance, we show the values for the ($n_{\varphi}=2$ and $n_r=2$) and the ($n_{\varphi}=2$ and $n_r=3$) resonance. Neglecting any other effects, the resonance widths in this table imply stars trapped in higher-order resonances should be more sensitive to perturbation by a given subhalo. However, when factoring in the true population of subhaloes at different Galactic radii, higher order resonances may reside in less perturbative environments.

\begin{table}
\centering
\renewcommand{\arraystretch}{1.2}
\begin{tabular}{cccc}
\hline
\multirow{2}{*}{Resonance ($n_{\varphi}=2$)} & \multicolumn{3}{c}{$\Delta I_{\rm half} \,[\Delta I_{\rm CR}]$}                  \\[0.8ex] \cline{2-4} 
                           & \multicolumn{1}{c}{$\Omega_{\rm p} = 35$} & \multicolumn{1}{c}{$40$} & $45$ \\ \hline\hline
Co-rotation ($n_r=0$)                & \multicolumn{1}{c}{$1.00$}          & \multicolumn{1}{c}{$1.00$}          & $1.00$          \\ \hline
Outer Lindblad ($n_r=1$)             & \multicolumn{1}{c}{$0.72$}          & \multicolumn{1}{c}{$0.75$}          & $0.76$          \\ \hline
1:1 resonance ($n_r=2$)      & \multicolumn{1}{c}{$0.56$}          & \multicolumn{1}{c}{$0.58$}          & $0.60$          \\ \hline
3:2 resonance ($n_r=3$)                 & \multicolumn{1}{c}{$0.46$}          & \multicolumn{1}{c}{$0.48$}          & $0.50$          \\ \hline
\end{tabular}
\caption{Comparison of higher order resonance widths $\Delta I_{\rm half}$ in units of the co-rotation resonance half-width, $\Delta I_{\rm CR}$ (for different bar speeds). The resonances are listed in order of increasing radius from the center of the galaxy. The units of $\Omega_{\rm p}$ are in km s$^{-1}$ kpc$^{-1}$.}
\label{tab:resonances}
\end{table}

\subsection{Perturbations to a resonant orbit}

Here we discuss the ways in which the slow action can be pushed beyond the separatrix. In this work, we operate primarily within the impulse approximation, whereby the slow angle of the resonant star fixed during an encounter. 

While we have established that $\Delta I_{\rm half}$ is the maximum width of the resonance island, this simple scalar is not the only criterion for determining ejection from resonance. There are three factors that determine a perturbers ability to cause a star to move beyond the separatrix and begin circulating. These are: a) the intrinsic properties of the perturber (i.e. mass, size, concentration), the geometry of the perturber (i.e. relative velocity direction and impact parameter direction), and the phase of the resonant star itself. The properties of the bar itself are an additional factor, which we neglect from this brief discussion.

In the case of dark matter subhaloes, the only perturber we consider in this paper, the first factor is obvious; all other properties things fixed, and neglecting the density profile, a more massive (or more concentrated) subhalo should provide a larger impulse. The second factor is slightly more nuanced. In the case of the co-rotation resonance, the slow action is primarily determined by the azimuthal action (or $z$-component of angular momentum) $L_z$. In the plane, for the impulse regime, the change in azimuthal action is $\Delta L_z = r \Delta v_{\varphi}$. Therefore the component of $\Delta \bm{v}$ that matter is the most is azimuthal direction $\Delta v_{\varphi}$. The most effective kicks are those with the velocity vector aligned with the azimuthal direction at the star's location. The third factor is even more subtle. A star librating around resonance can be pushed beyond the separatrix by a perturber that imparts a smaller change in slow action than $\Delta I_{\rm half}$. Since the slow angle librates, the distance to the separatrix changes throughout the star's orbit. In brief, the resonance phase controls how large of a change in velocity $\Delta v_{\rm \varphi}$ (or equivalently $\Delta L_z$) is required to eject the star from resonance.

In summary: the intrinsic properties of subhalo controls the magnitude of the change of velocity $|\Delta \bm{v}|$, the geometry of the encounter controls how much of that velocity change lands in the azimuthal direction $\Delta v_{\varphi}$, and the resonance phase controls the required amount of $\Delta v_{\varphi}$. When exploring the intrinsic properties of subhaloes that are able to eject stars from resonant orbits, we must therefore consider the geometry and phase of the encounter carefully. It is most informative to explore cases for which the geometry and phase are either maximally or minimally impactful to get bounds on the intrinsic subhalo properties that cause a star to be ejected from resonance.

\section{Single subhalo impulse}\label{sec:single_subhalo}

\begin{figure}
\centering
\begin{tikzpicture}[x=1cm,y=1cm]
  \fill[black!15] (-1,-2.5) ellipse [x radius=1.5, y radius=0.4];
  \draw[thick, black!50]    (-1,-2.5) ellipse [x radius=1.5, y radius=0.4];

  \draw[thick, orange!50, dashed]
    (-3.5,-1.2)
      .. controls (-2,0.3) and (0,0.3) .. (1.5,-1.2)
      .. controls (0,-0.3) and (-2,-0.3) .. (-3.5,-1.2);

    \draw[-stealth, thick]
        (1.3,-2.1) arc[start angle=70, end angle=-70, radius=0.4];

    \node[right] at (1.7,-2.6) {$\Omega_{\rm p}$};

  \coordinate (OrbitCenter) at (-1.00,-4.00);
  \def\OrbitR{5.3125}
  \def\angA{140.00} 
  \def\angB{40.00}  
  \def\angM{90.00}  

  \draw[thick, black!50, dashed]
    (OrbitCenter) ++(\angA:\OrbitR)
    arc[start angle=\angA, end angle=\angB, radius=\OrbitR];

    \node[
      star,
      star points=5,
      star point ratio=2.25,
      fill=orange,
      draw=none,
      inner sep=2pt
    ] at (-1.0,-0.05) {};

  \coordinate (Subhalo) at ($(OrbitCenter)+(\angM:\OrbitR)$);
  \fill (Subhalo) circle (5.0pt) node[above right, xshift=5pt] {$M$};

  \draw[-stealth, ultra thick, orange]
    (-1, -0.05) -- (-2, -0.05);

  \draw[-stealth, ultra thick, black]
    (Subhalo) -- ++(-1, 0.00);
    \node[anchor=north east] 
      at (current bounding box.north east) 
      {\textbf{Co-rotating frame}};

    \path (-5,2) rectangle (3,-3.5);

    \draw[black!30, line width=0.6pt]
      ([xshift=-3pt,yshift=-3pt]current bounding box.south west)
      rectangle
      ([xshift=3pt,yshift=3pt]current bounding box.north east);
\end{tikzpicture}

\vspace{4mm}

\begin{tikzpicture}[x=1cm,y=1cm]

    \draw[black!60, dashed]
      (-1,-0.05) -- (-1,2)
      node[midway, left] {$\bm{b}$};

    \draw[black!60, dashed]
      (-1,-0.05) -- (2,2)
      node[midway, below, right, yshift=-7pt, xshift=-6pt] {$\bm{r}(t)=\bm{b}+\bm{u}(t-t_{\rm enc})$};

    \path (-5,3) rectangle (3,-1);

    \fill[black!10] (2,2) circle (5.0pt);

    \draw[thick, black!50, dashed] (-5,2) -- (3,2);

    \node[
      star,
      star points=5,
      star point ratio=2.25,
      fill=orange,
      draw=none,
      inner sep=2pt
    ] at (-1.0,-0.05) {};

    \fill (-1,2) circle (5.0pt);

    \draw[-stealth, ultra thick, black]
        (-1, 2) -- (-2, 2)
         node[above] {$\bm{u} = \bm{w} - \bm{v}$};

    \node[anchor=north east] 
      at (current bounding box.north east) 
      {\textbf{Star's rest frame}};

    \draw[black!30, line width=0.6pt]
      ([xshift=-3pt,yshift=-3pt]current bounding box.south west)
      rectangle
      ([xshift=3pt,yshift=3pt]current bounding box.north east);

\end{tikzpicture}
\caption{Schematic diagram of subhalo flyby of a star in resonance with the galactic bar in the impulse approximation, where the subhalo travels on a straight line orbit in the Galactic rest frame. The top panel shows the co-rotating frame of the bar (with pattern speed $\Omega_{\rm p}$), whereas the bottom panel shows the instaneous rest frame of the star. The grey ellipse in the top panel represents the bar. The subhalo's orbit is represented by a dashed black line and circle, and the star's co-rotation resonant orbit is represented by a dashed orange line and star. In the star's rest frame, the subhalo passes with instantaneous velocity $\bm{u} = \bm{w} - \bm{v}$, where $\bm{w}$ and $\bm{v}$ are the instantaneous velocities of the subhalo and star respectively. The vector $\bm{r}(t)$ defines the local linearized trajectory, with impact parameter vector $\bm{b}$.}
\label{fig:impulse_diagram}
\end{figure}
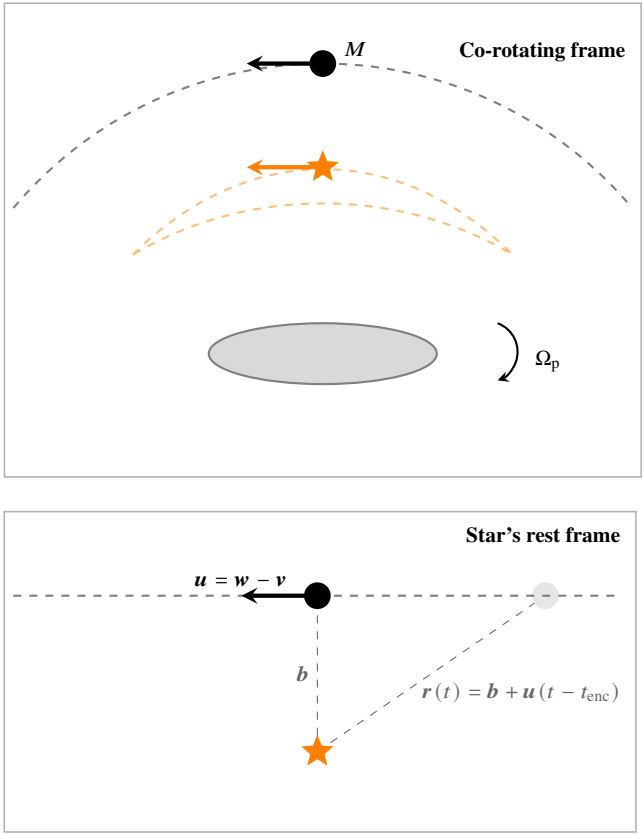

Now that we have established the criterion for ejecting the parent orbit from resonance, we can compute the change in slow action imparted by a subhalo $\Delta I_{\rm sh}$. A star will be ejected from resonance when 
\begin{equation}
    \Delta I_{\rm sh} > \Delta I_{\rm half}.
\end{equation}
This condition is subject to the geometry of the encounter and the libration phase of the star. In this section we use the impulse approximation to find an analytic expression for $\Delta I_{\rm sh}$. We subsequently compare this analytic impulse model with a simulation of a test particle in a barred potential.

\subsection{Analytic impulse model}

If we assume that the forces of the subhalo act on timescales much smaller than the orbital time scale of the resonant star, then we can apply the impulse approximation,
\begin{equation}\label{eq:impulse}
    \Delta \bm{v} = \int_{-\infty}^{\infty}\bm{a}(t)dt,
\end{equation}
where the change in the star's velocity $\Delta \bm{v}$ is simply the integral over the gravitational force on the star from the subhalo $\bm{a} = -\nabla \phi$. In the impulse approximation, we the treat the star as stationary and the subhalo as moving on a straight line.

Throughout this work we model any subhaloes by a Plummer sphere of mass $M$ with scale radius $r_{\rm s}$. This is primarily done for analytic tractability, but also follows subhalo treatment in \citet[][]{erkal2015properties, erkal2016number}. The form of the potential for a Plummer sphere is
\begin{equation}\label{eq:plummer}
    \Phi_P(r) = -\frac{GM}{\sqrt{r^2 + r_{\rm s}^2}}.
\end{equation}
In the instantaneous rest frame of the of the star, the relative velocity of subhalo and star is
\begin{equation}
    \bm{u} = \bm{w} - \bm{v}, 
\end{equation}
where $\bm{v}$ is the velocity of the star at the moment of the encounter, and $\bm{w}$ is the subhalo velocity is in the Galactic rest frame. As shorthand, we write the magnitude of the relative velocity as $u = |\bm{u}|$. While stars on resonances can be found throughout the halo at a variety of orientations, we simplify this work by assuming the resonant star's orbit is confined to the $x-y$ plane. At the time of interaction with the subhalo we approximate the star as moving in a straight line.

Each encounter is parametrised by an impact parameter $\bm{b}$, which is the separation of the star and the subhalo at the moment of closest approach. We can write the relative position of the subhalo with respect to the star as 
\begin{equation}
    \bm{r}(t) = \bm{b} + \bm{u}(t - t_{\rm enc}).
\end{equation}
This defines $t=t_{\rm enc}$ as the time of closest approach, i.e. $\bm{b} = \bm{r}(t_{\rm enc})$. By minimising the distance (i.e. the definition of closest approach) we find that $\bm{b}\cdot\bm{u} = 0$, meaning that $\bm{b}$ lies in the 2-dimensional plane perpendicular to $\bm{u}$. We denote the magnitude of the impact parameter $b\equiv|\bm{b}|$. In Fig.~\ref{fig:impulse_diagram} we illustrate the subhalo's flyby of the resonant star as a visual guide. 

Using the equation for the Plummer sphere, we know the gravitational force exerted on the star is given by
\begin{equation}\label{eq:force_plummer}
    \bm{a}(t) = -GM\frac{\bm{r}(t)}{(\bm{r}(t)^2 + r_{\rm s}^2)^{3/2}}.
\end{equation}
By integrating Eq.~\ref{eq:impulse}, we get an expression for the change in velocity of the resonant star as a result of the subhalo fly-by:
\begin{equation}
    \Delta \bm{v} = \frac{2GM}{u} \frac{\bm{b}}{(b^2 + r_{\rm s}^2)}.
\end{equation}
This final velocity impulse can be approximately converted to an action impulse by considering the relationship between the slow action and the spherical coordinate actions. In the case of the slow action, we only need to consider the azimuthal action (i.e. the z-component angular momentum). Namely, $I = L_z / n_{\varphi}$ and thus
\begin{equation}
    \Delta I = \Delta L_z / n_{\varphi}.
\end{equation}
We recognise that, from the definition of actions in Eq.~\ref{definition_action}, we can relate the magnitude of impulse velocity to the impulse in angular momentum $L_z$ by
\begin{equation}
     \Delta L_z =(\bm{r} \times \Delta\bm{v})_z = r \, \Delta v_{\varphi},
\end{equation}
where $\Delta v_{\varphi} = \Delta \bm{v}\cdot\bm{e}_{\varphi}$. Here, $\bm{e}_{\varphi}$ is the azimuthal unit vector pointing perpendicular to the radial direction from the Galactic centre. Plugging in for $\Delta v_{\varphi}$, $b_{\varphi} = \bm{b}\cdot\bm{e}_{\varphi}$, and $\Delta L_z$, we get
\begin{equation}
     \Delta I_{\rm sh}= \frac{r}{n_{\varphi}} \frac{2GM}{u} \frac{b_{\varphi}}{(b^2 + r_{\rm s}^2)}
\end{equation}
as the expression for the action impulse for an encounter between a subhalo of mass $M$ for a star on an orbit with some characteristic radius radius $r$. We typically take $r$ as the radius of the resonance $r_{\rm res}$ as given by Eq.~\ref{eq:res_radius}. 

\subsubsection{Individual subhalo properties}

Now we know how the intrinsic subhalo properties (i.e. mass $M$ and scale radius $r_s$) determine a change in slow action $\Delta I$ (or equivalently a change in azimuthal action $L_z$), we need to decide how to set these properties to be cosmologically appropriate. This requires adopting a prescription that relates the internal structure of subhaloes to their mass, which in turn depends on the assumed dark matter model. While CDM sets a certain relationship, alternative models like SIDM may allow for more concentrated subhaloes. We therefore explore a few different radius-mass $r_s-M$ relationships. 

The scale radius of a Plummer sphere can be found from its mass by \citep[][]{diemand2008clumps, erkal2016number}
\begin{equation}\label{eq:scale_radius}
    r_{\rm s}(M) = 1.62 {\,\rm kpc} \left(\frac{M}{10^8 {\rm M}_{\odot}}\right)^{1/2},
\end{equation}
which approximately reproduces the effective sizes of CDM subhaloes when represented as Plummer spheres. 

\subsection{Computational model}\label{sec:simulations}

To complement the analytic model and validate our assumptions, we conduct a series of test particle simulations that emulate a dark matter subhalo impacting a star in co-rotation resonance with a Milky Way-like bar potential.  Specifically, we conduct simulations of a \textit{single} subhalo passing by a star in co-rotation resonance, and compare with this what is expected using impulse approximation. All of the simulations are produced using the Agama library \citep[][]{vasiliev2019agama}.

Our main aim is to use these simulations to get a better understanding of whether our analytical model either under predicts and or over predicts the change in action imparted by a subhalo. Moreover, we a want to simply examine whether or not a single subhalo can impart enough change in action to eject a star from resonance. The result of this experiment will motivate the exploration of the impact of a population of subhaloes.

\subsubsection{Simulation components}

Here we describe the individual components of the simulation, which consist of a host potential, a test particle representing a star in resonance with a bar, and a dark matter subhalo on a pre-determined trajectory.

\begin{figure}
    \centering
    \includegraphics[width=0.8\columnwidth]{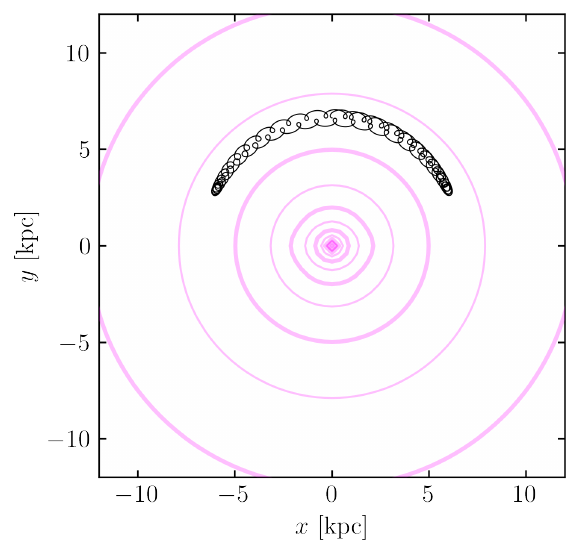}
    \caption{An example orbit (black line) in co-rotation resonance with the bar. The bar's major axis is aligned with the $x$-axis, and the equipotential lines are shown in magenta.}
    \label{fig:resonance}
\end{figure}

The potential of the host galaxy is kept simple, and consists of just two components: a cored logarithmic halo potential (as described by Eq.~\ref{logarithmic}) with $v_0 = 230$ km/s and $r_{\rm core} = 0.1$ kpc, and a Ferrers bar potential with parameters chosen so that the force along the major axis of the bar component alone matches that of the analytic model in Eq~\ref{eq:bar_potential} with pattern speed $\Omega_{\rm p} = 35$ km s$^{-1}$ kpc $^{-1}$, strength $A=0.02$, and scale length $r_{\rm b} = 1.6$ kpc. Matching the major-axis force ensures that the Ferrers bar reproduces the dynamical effect of the analytic perturbation, while introducing a self-consistent mass distribution for the bar. In Fig.~\ref{fig:resonance} we present a contour plot showing the equipotential lines (in magenta) of the entire potential used in the test particle simulation, alongside a co-rotation resonant orbit (in black). This figure is in the co-rotating frame. In this plot, the bar's major axis is aligned with the $x$-axis, and we can see that the trajectory of the co-rotation resonance is confined to one side of the bar, as expected. 

The resonant star's orbit is generated by satisfying the resonance condition described by Eq.~\ref{res_condition_omega}, and then setting it on an initial position $\bm{x}_* = (0, \, r_{\rm res}, \, 0)$ and velocity $\bm{v}(t=0) = -(0, \, v_{\rm circ}(r_{\rm res})+\varepsilon,\,0)$, where $v_{\rm circ}(r)$ is the circular velocity at in the logarithmic potential and $\varepsilon$ is some small initial offset velocity. Evident from the initial conditions, the star is confined to the $x-y$ plane. In Fig.~\ref{fig:resonance}, the example resonant orbit is chosen with $\varepsilon = 10$ km/s. In every simulation in this work, we integrate the orbit for a total of $8$ Gyrs.

The subhalo potential is modelled as a Plummer sphere (Eq.~\ref{eq:plummer}), moving on a pre-determined straight-line trajectory through the host potential. We work in the instantaneous rest frame of the star at the time of closest approach, $t=t_{\rm enc}$. Defining the relative separation vector
\begin{equation}
    \bm{r}(t) \equiv \bm{x}_{\rm sh}(t) - \bm{x}_\star(t_{\rm enc}),
\end{equation}
the encounter geometry is specified by $\bm{r}(t_{\rm enc}) = \bm{b}$, where $\bm{b}$ is the impact parameter vector. The subsequent motion of the subhalo is then
\begin{equation}
    \bm{r}(t) = \bm{b} + \bm{u}\,(t - t_{\rm enc}),
\end{equation}
where $\bm{u} = \bm{w} - \bm{v}$ is the constant relative velocity vector. By construction, $t_{\rm enc}$ corresponds to the time of closest approach.

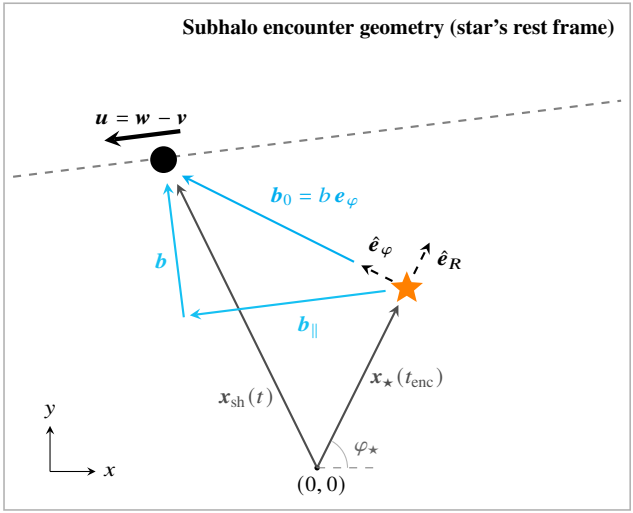
\begin{figure}
    \centering
\begin{tikzpicture}[x=1cm,y=1cm]

\def\xstar{0.18}
\def\ystar{0.36}

\def\xsh{-3.0}
\def\ysh{1.95}

\fill (-1.0,-2.0) circle (1pt);
\node[below left] at (-0.5,-2.0) {$(0,0)$};

\draw[-stealth, dashed, thick]
    (\xstar,\ystar) -- ++(0.3,0.6)
    node[below right] {$\hat{\bm{e}}_R$};

\draw[-stealth, dashed, thick]
    (\xstar,\ystar) -- ++(-0.6,0.3)
    node[above right] {$\hat{\bm{e}}_\varphi$};

\draw[-stealth, thick, black!70, shorten >=7pt]
    (-1.0,-2.0) -- (\xstar,\ystar)
    node[midway, right] {$\bm{x}_\star(t_{\rm enc})$};

\draw[-stealth, thick, black!70, shorten >=7pt]
    (-1.0,-2.0) -- (-2.95,1.95)
    node[midway, left, yshift=-30pt, xshift=15pt] {$\bm{x}_{\rm sh}(t)$};

\draw[-stealth, thick, cyan, shorten >=7pt, shorten <=3pt]
    ({\xstar-0.6},{\ystar+0.3}) -- (-3.0,1.95)
    node[midway, above right] {$\bm{b}_0 = b \, \bm{e}_\varphi$};

\def\k{-0.3075}
\pgfmathsetmacro{\xb}{\xsh + (-1)*\k}   
\pgfmathsetmacro{\yb}{\ysh + (8)*\k}    

\pgfmathsetmacro{\yparL}{\ystar + (-4-\xstar)*0.125} 
\draw[cyan!70, -stealth, thick, shorten <=8pt, shorten >=38pt]
    (\xstar,\ystar) -- (-4,\yparL)
    node[midway, below left, xshift=30pt] {$\bm{b}_{\parallel}$};

\draw[cyan!70, -stealth, thick, shorten >=6pt, shorten <=14pt]
    (\xb,\yb) -- (\xsh,\ysh)
    node[midway, left] {$\bm{b}$};

\path (-5,4) rectangle (3,-1);

\draw[thick, black!50, dashed] (-5,1.825) -- (3,2.825);

\fill (-3.02,2.05) circle (5.0pt);

\draw[-stealth, ultra thick, black]
    (-2.8,2.425) -- (-3.8,2.3)
    node[midway, above, yshift=1] {$\bm{u} = \bm{w} - \bm{v}$};

\node[
  star,
  star points=5,
  star point ratio=2.25,
  fill=orange,
  draw=none,
  inner sep=2pt
] at (\xstar,\ystar) {};

\draw[dashed, black!50]
    (-1.0,-2.0) -- (-0.2,-2.0);

\draw[black!30]
    (-0.6,-2.0) arc[start angle=0, end angle=60, radius=0.4];

\node[black!60] at (-0.35,-1.75) {$\varphi_\star$};

\begin{scope}[shift={(current bounding box.south west)}, xshift=0.5cm, yshift=0.4cm]
    \draw[-stealth, thin] (0,0) -- (0.6,0) node[right] {$x$};
    \draw[-stealth, thin] (0,0) -- (0,0.6) node[above] {$y$};
\end{scope}

\node[anchor=north east]
  at (current bounding box.north east)
  {\textbf{Subhalo encounter geometry (star's rest frame)}};

\draw[black!30, line width=0.6pt]
  ([xshift=-3pt,yshift=-3pt]current bounding box.south west)
  rectangle
  ([xshift=3pt,yshift=3pt]current bounding box.north east);

\end{tikzpicture}
    \caption{Geometry of a subhalo encounter in the star's instantaneous rest frame at time $t$. The star is located at $\bm{x}_\star(t_{\rm enc})$ and is described by the local cylindrical basis $\hat{\bm e}_R$ and $\hat{\bm e}_\varphi$. The subhalo moves on a linear trajectory with relative velocity $\bm{u} = \bm{w} - \bm{v}$. The trial impact parameter $\bm{b}_0 = b\,\hat{\bm e}_\varphi$ is decomposed into components parallel and perpendicular to the trajectory, $\bm{b}_{\parallel}$ and $\bm{b}$, respectively. This geometry defines the impulse applied to the star in the simulations.}
    \label{fig:sim_geometry}
\end{figure}

We consider two classes of trajectories for the subhalo: vertical (perpendicular) encounters, for which $\bm{w} = w_{0}\,\bm{e}_z$, and in-plane (parallel) encounters, for which $\bm{w}$ lies within the disc plane and is aligned with the instantaneous direction of motion of the star at $t_{\rm enc}$. In both cases, the impact parameter vector at closest approach satisfies the standard condition $\bm{b} \cdot \bm{u} = 0$. Changes to the slow action are directly related to changes in $L_z$, which are driven by tangential velocity kicks. To maximise this effect, we therefore choose a preferred impact direction aligned with the star’s instantaneous azimuthal unit vector $\hat{\bm{e}}_{\varphi} = (-\sin\varphi_*,\, \cos\varphi_*,\, 0) $ where $\varphi_*$ is the star’s azimuth at the encounter. We first define a trial vector, centred at the star's location
\begin{equation}
    \bm{b}_0 = b \, \hat{\bm{e}}_{\varphi},
\end{equation}
and then enforce the closest-approach condition by projecting this vector into the plane perpendicular to the relative velocity,
\begin{equation}
    \bm{b} 
    = 
    \bm{b}_0 
    - 
    (\bm{b}_0 \cdot \hat{\bm{u}})\,\hat{\bm{u}},
\end{equation}
where $\hat{\bm{u}} = (\bm{w} - \bm{v})/|\bm{w} - \bm{v}|$ is the unit vector in the direction of the relative velocity at $t_{\rm enc}$. This encounter geometry is presented diagrammatically in Fig.~\ref{fig:sim_geometry}.

\subsection{Comparison of computational and analytic model}

Here we conduct a suite of simulations to better contextualise the analytic model, and quantify its limitations. The components described above are combined in a routine whereby we loop over different subhalo properties (mass, relative velocity, impact parameter) and compare the result with what is predicted by the analytic model. While we vary the subhalo properties, the bar properties are fixed across each simulation. Therefore, this is not a comprehensive comparison, but aims to provide a sense of where the analytic model may break down for a Milky-Way like bar. This comparison is only done for the co-rotation resonance, as this is only where the analytic model is applied. For a given triplet of subhalo parameters $(M,b,\bm{w})$, we additionally loop over a variety of resonant star slow angles at the time of subhalo closest approach, and subsequently take the mean. The detailed steps are as follows:

\begin{enumerate}
    \item Integrate the resonant orbit until many libration cycles are sampled;
    \item Identify the slow angle, $\upphi$;
    \item Select a set of slow-angle $\{\upphi_i\}$ values uniformly across $[-\pi, \pi]$;
    \item For each $\upphi_i$, determine the time $t_i$ at which the orbit attains that slow angle;
    \item Position the subhalo so that its closest approach occurs at $t=t_{\rm enc}$.
\end{enumerate}

This produces a collection of encounters that uniformly sample the slow angle of the resonant oscillation. The final ``impact'' for a given subhalo property triplet $(M,b,\bm{w})$ is taken as the mean over all sampled slow angles.

How we define and compare the ``impact'' between the test particle simulation and the analytic approximation is not a straightforward question given that neither velocity nor $z$-component angular momentum are integrals of motion in a barred potential. This makes it non-trivial to choose over what timescale after subhalo passage to compare the simulated $\Delta v$ or $\Delta L_z$ with the impulse approximated case. We must therefore compare the simulations at a few different times.

\subsubsection{Comparison method}

Since the slow action $I$ couples directly to $L_z$, we begin by calculating the change in $L_z$ for different subhalo properties. Using the change in $z$-component angular momentum $\Delta L_z$, we can compare the toy analytic model (labelled with superscript `toy') to the test particle simulation (labelled with superscript `sim'). We run a grid of simulations across mass, relative velocity and impact parameter to see in what ranges of these parameters the analytic model over predicts or under predicts. Specifically, we run a $6 \times 40 \times 40$ grid for $(M,b,u)$, where the ranges of these parameters are $M=[10^5,10^{10}]$~M$_{\odot}$, $b=[0,10]$~kpc, and $u = [0,500]$~km/s. All the subhaloes in this simulation grid have standard CDM mass-radius relationships (Eq.~\ref{eq:scale_radius}). We calculate the difference of $\Delta L_z$ in the case of the analytical model and the simulation:
\begin{equation}
    \delta_{ \Delta L_z} = \frac{\Delta L_z^{\rm (toy)} - \Delta L_z^{\rm (sim)}}{|\Delta L_z^{\rm (toy)}| + |\Delta L_z^{\rm (sim)}|}.
\end{equation}
This parameter lets us see where the analytic model for the impulse approximation under predicts and where it overpredicts compared to the simulation. 

For each mass bin we calculate the coefficient of determination, $R^2$, treating the simulated impulse $\Delta L_z^{\rm (sim)}$ as the true values and $\Delta L_z^{(\rm toy)}$ as the predicted values. Since we are comparing the impulse approximation analytic model with a more realistic simulation, it is unclear as to when precisely after the subhalo--star closest $t_0$ the two methods should match. More precisely, the slow action $L_z$ is not an integral of motion, and therefore the change in slow action for the simulation $\Delta L_z^{\rm (sim)}$ is not constant after the subhalo passage. We must therefore decide how long after the subhalo passage $\Delta t$ to compare the simulation and analytic model. That is to say, if the subhalo has closest approach with the star at time $t=t_{\rm enc}$,  we calculate $R^2$ at time $t=t_{\rm enc} + \Delta t$. We choose to calculate the $R^2$ value for a variety of $\Delta t$ values and compare them all. 

In the case where the simulation perfectly matches the analytic model, $R^2 = 1$. A value of $R^2 = 0$ suggests that the toy model simply matches the average value of the simulation, and $R^2 < 0$ means that the simulation is actively worse than just guessing the mean. In addition to calculating $R^2$ for each mass bin, we calculate it for all mass bins together. Calculation of $R^2$ is done using the \texttt{sklearn} function \texttt{r2\_score}, with flattened arrays. 

One may ask why we do not use the Jacobi integral $E_J$, which is an integral of motion in the bar potential, as our simple comparison metric. Although the Jacobi integral is constant along a librating orbit in the rotating frame, the change in the Jacobi integral $\Delta E_J$ induced by an impulsive encounter depends on the libration phase at which the encounter occurs. As a result, it does not provide a phase-independent measure of resonant diffusion or escape.

\subsubsection{Comparison results}\label{sec:comparison_results}

Here we use the metrics discussed in the previous sub-section to assess the how similar the analytic model is to the simulation. 

In Table.~\ref{tab:r2_scores} we show the $R^2$ value for all the simulation mass bins (individually and all-together), for a few different values of $dt$. It is to be expected that the comparison is not reasonable at the exact moment of closest approach, $\Delta t=0.000$ Gyr, or long after the subhalo passage $\Delta t \gtrsim 1$~Gyr. We choose to compare the simulations at the times $\Delta t~=~\{0.000, 0.010, 0.025, 0.050, 0.100, 0.200\}$ Gyr. Table.~\ref{tab:r2_scores} illustrates that the impulse approximation formula best matches the simulation $\Delta t = 0.025$ Gyr after the closest approach of the subhalo, and for values of $M<10^9$ M$_{\odot}$. For all masses, the $R^2$ value is $0.71$. This $R^2$ value means that 71\% of the variation in the outcome is explained by the analytic model's variables. This table demonstrates that the analytic model is not a reasonable description for masses above $M=10^9$ M$_{\odot}$. For masses $M=M^5-10^7$ M$_{\odot}$, the analytic model matches the simulations to $R^2>0.66$ for all $\Delta t$ other than $\Delta t=0.000$ Gyr and $\Delta t=0.200$ Gyr. 

\begin{figure}
    \centering
    \includegraphics[width=\columnwidth]{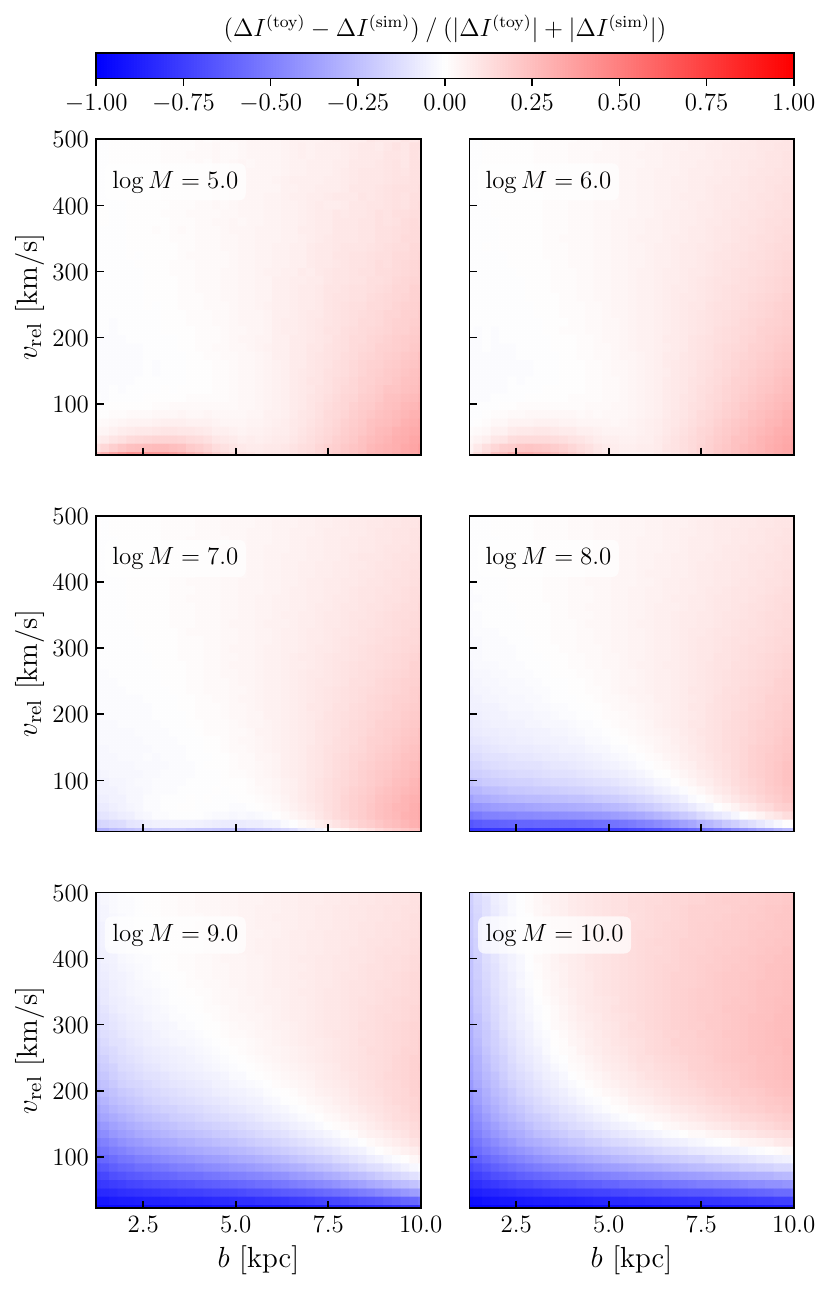}
    \caption{The difference between the change in azimuthal action $L_z$ predicted from the toy model and the simulation. This comparison is only for an impact by one subhalo. We show normalised $\Delta L_z$ in $v_{\rm rel}-b$ space for various subhalo masses. Blue colour indicates an under estimate by the toy analytic model relative to the simulation, whereas red colour indicates an over estimate.}
    \label{fig:comparison_slow}
\end{figure}

In Fig.~\ref{fig:comparison_slow} we present the values of $\delta_{\Delta I}$ to inspect for what regions of $(M,b,v)$ space the analytic model underpredicts (blue regions) and where it overpredicts (red regions). Specifically, we present the case of $\Delta t=0.025$ Gyr, as it is at this time that the analytic model best matches the simulation. While this is the best matching, we see many regions of $(b,v)$ parameter space where the analytic model breaks down. As expected, very low relative velocities do not match the impulse approximation. We find that that the analytic model is much more underpredictive (see the location of the blue regions) when the subhalo has a high mass, and when it is at low relative velocity. In contrast, the analytic model is much more overpredicted (see the location of the red regions) at high masses.

\begin{table}
\centering
\renewcommand{\arraystretch}{1.2}
\resizebox{\linewidth}{!}{
\begin{tabular}{ccccccc}
\hline
\multirow{2}{*}{Mass [M$_{\odot}$]} & \multicolumn{6}{c}{$R^2$ score} \\
 \cline{2-7}
 & $\Delta t=0.000$ & 0.010 & 0.025 & 0.050 & 0.100 & 0.200 \\
\hline
\hline
$10^{5}$  & -3.45  & 0.83  & 0.94  & 0.76  & 0.67 & -0.18 \\
$10^{6}$  & -2.91  & 0.79  & 0.97  & 0.79  & 0.70 & -0.13 \\
$10^{7}$  & -1.80  & 0.66  & 0.97  & 0.81  & 0.70 & -0.46 \\
$10^{8}$  & -1.33  & 0.26  & 0.62  & 0.66  & 0.36 & -3.86 \\
$10^{9}$  & -4.17  & -2.88 & -3.12 & 0.10  & -1.21 & -9.74 \\
$10^{10}$ & -50.87 & -24.50 & -4.19 & -2.52 & -11.22 & -26.84 \\
\hline
\rowcolor{gray!20}
$10^{5} - 10^{10}$       & -0.08  & 0.51  & 0.71  & 0.43  & -1.59 & -8.28 \\
\hline
\end{tabular}
}
\caption{$R^2$ score (i.e. coefficient of determination) for different subhalo masses $M$ and time intervals after subhalo nearest approach $\Delta t$. The time interval $\Delta t$ is measured in Gyrs.}
\label{tab:r2_scores}
\end{table}

\subsection{Demonstration and summary of single subhalo impact}

From the simulation suite described earlier, we choose some illustrative examples to explore the time evolution of the star's orbit after perturbation by the subhalo. We both visually inspect the way an orbit changes in configuration space $(x,y,z)$, and determine how the orbital properties $(L_z, E_J)$ of the star change over time. Moreover, we demonstrate that (using the $r_s-M$ relationship in Eq.~\ref{eq:scale_radius}) only a very limited part of parameter space for single subhaloes has the ability to eject a star from resonance. 

\begin{figure*}
    \centering
    \includegraphics[width=\textwidth]{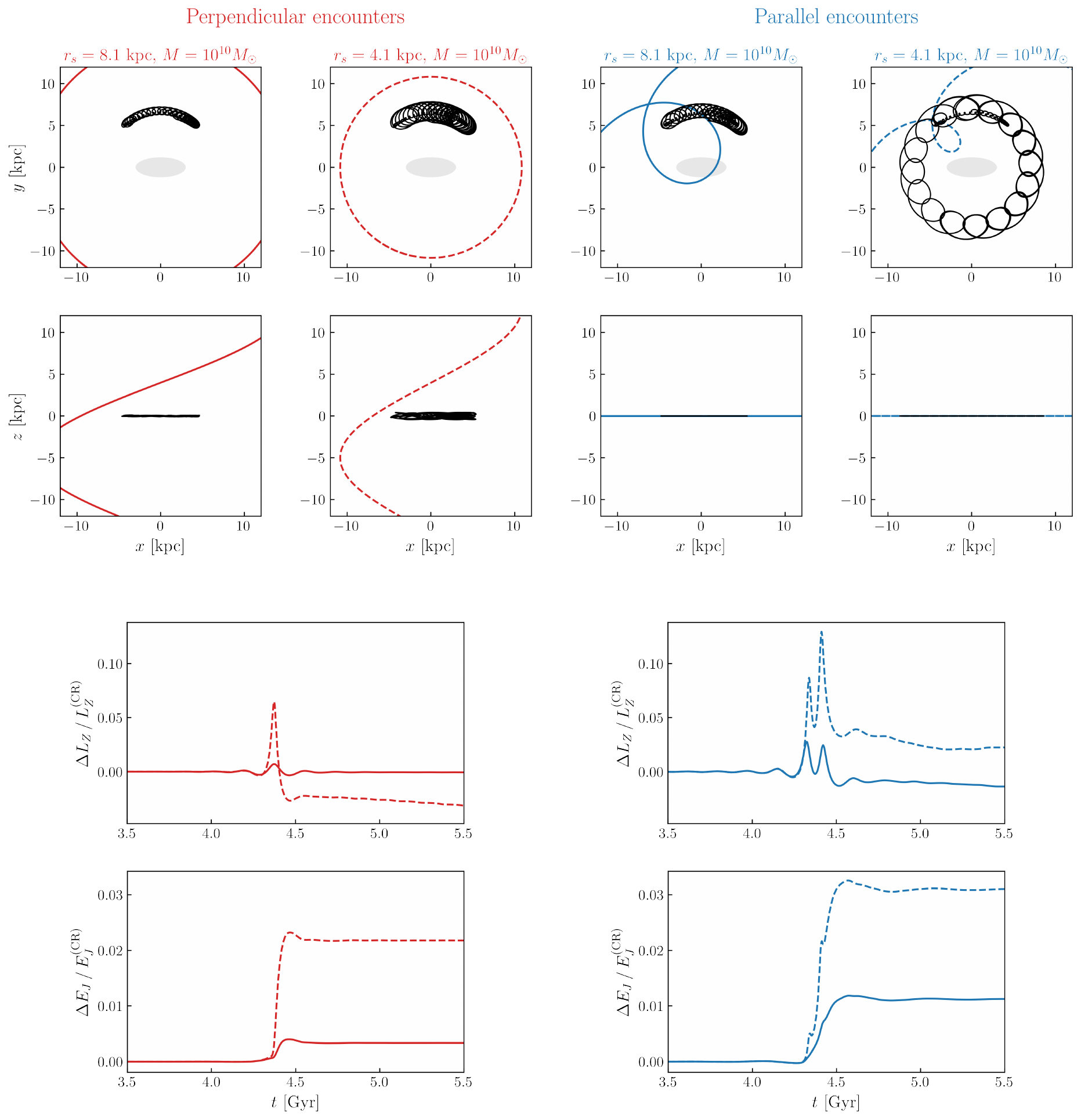}
    \caption{Time evolution of test particle resonant star orbits after passage by perturbing subhaloes. The left column (red orbit) shows the impact of a subhalo on a trajectory perpendicular to the star, while the right column (blue orbit) shows the impact of a subhalo on a trajectory parallel to the star. In each case, the impact parameter $b$ is equal to the scale radius $r_s$. For each encounter geometry type we show a subhalo with two different $r_s-M$ relationships. The upper eight panels illustrate various ways in which the resonant orbit can be changed. Note how the most concentrated subhalo in the parallel encounter causes the star to circulate freely. The lower four panels demonstrate the time evolution of the Jacobi integral $E_J$ and $z$-component angular momentum $L_z$. In all examples, the $b=r_s$, and the subhalo velocity is $ |\bm{w}|=200$ km/s.}
    \label{fig:orbits}
\end{figure*}

\begin{figure}
    \centering
    \includegraphics[width=0.9\linewidth]{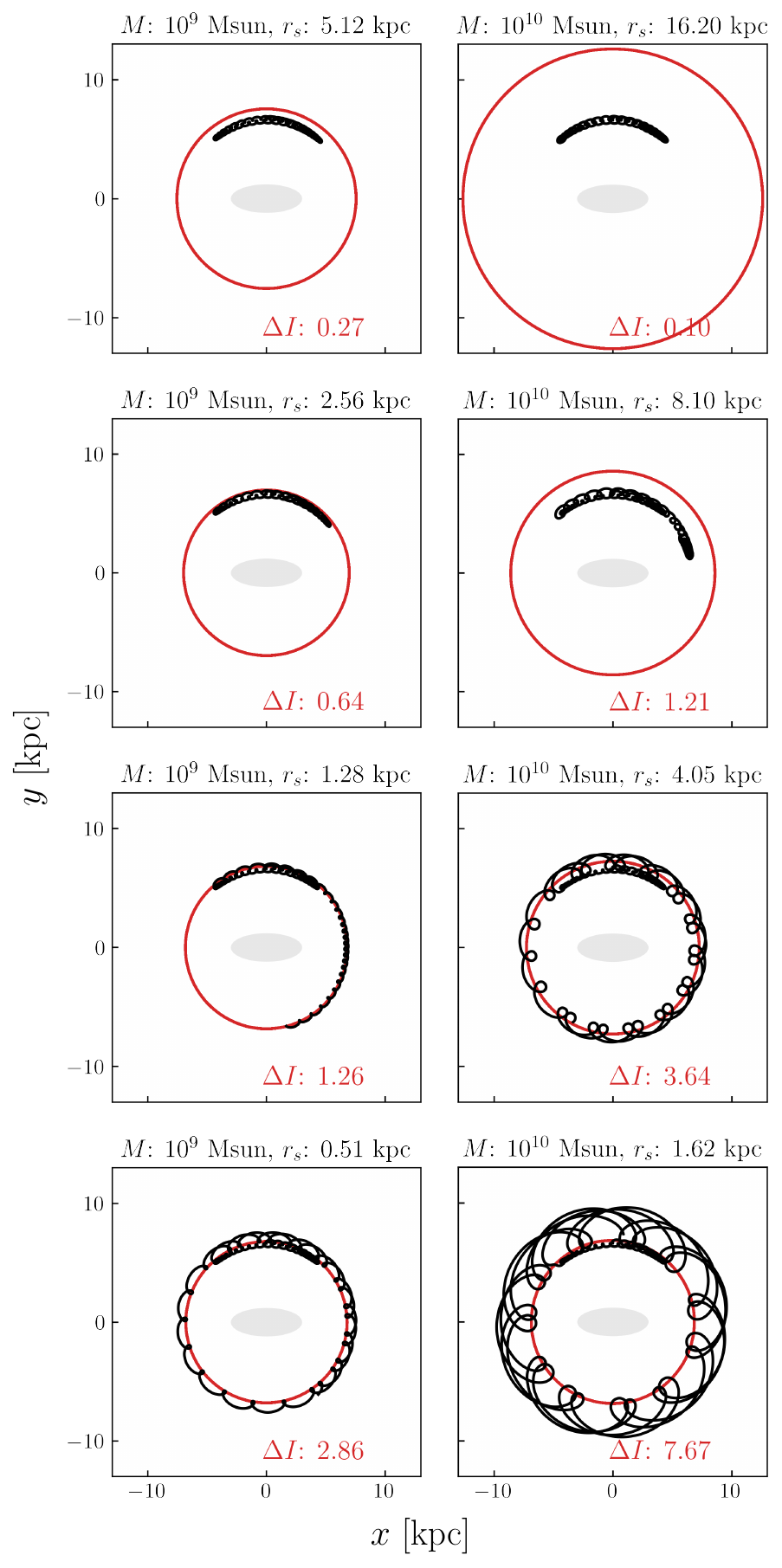}
    \caption{Example outcomes of single-subhalo fly-bys for various subhalo masses $M$ and scale radii $r_s$, all on vertical (perpendicular) trajectories. Each panel shows the orbit of a test particle star initially trapped in the co-rotation resonance (black) after interaction with a subhalo (red). The values of $\Delta I$ indicate the resulting change in slow action in units of the separatrix width. As subhalo mass increases (or scale radius decreases at fixed mass), the perturbation grows from mild deformation of the librating orbit to complete ejection from resonance, with the star transitioning to circulating motion. In these examples, the impact parameter $b=r_s$, and the subhalo vertical velocity is $w_0 = 200$ km/s.}
    \label{fig:orbits_more}
\end{figure}

Given that the encounter geometry has a large impact in determining whether or not a subhalo is able to eject a star from resonance, in Fig.~\ref{fig:orbits} we present two encounters with distinct subhalo trajectories: a) vertical (perpendicular), and b) in-plane (parallel). In each case the impact parameter is fixed to $b=r_s$, and the subhalo velocity is $|\bm{w}| = 200$ km/s. The upper panels show the orbital evolution in configuration space, while the lower panels show the time evolution of the Jacobi Integral $E_J$ and the azimuthal angular momentum $L_z$. In Figure~\ref{fig:orbits_more} we present a few more examples orbits of varying scale radius and mas, labelled by their change in slow action $\Delta I$ (in units of total resonance separatrix width $2\Delta I_{\rm half}$). Based on the results in Sec.~\ref{sec:comparison_results}, we evaluate the slow action at at $t = t_{\rm enc} + 0.025\,{\rm Gyr}$. In both Fig.~\ref{fig:orbits} and Fig.~\ref{fig:orbits_more}, the scale radii are chosen using Eq.~\ref{eq:scale_radius}, multiplied by some factor $f_s$. In Fig.~\ref{fig:orbits} we present $f_s = \left[0.5,1.0 \right]$ and in Fig.~\ref{fig:orbits_more} we present $f_s = \left[0.1, 0.25, 0.5, 1.0 \right]$. \\

There are three broad regimes of impact: 
\begin{enumerate}
    \item $\Delta I \ll 1$ -- the star can experience a weak perturbation, where it is guaranteed to remain within the separatrix boundary. There is effectively no visible change to the orbit;
    \item $\Delta I \approx 1$ -- the star can experience a marginal perturbation, where its ejection is heavily phase-dependent. The orbit may begin circulating, or the star's radial amplitude may increase but remain in co-rotation;
    \item $\Delta I \gg 1$ -- the star can experience a large perturbation, where ejection from resonance is guaranteed. The orbit will visibly circulate.
\end{enumerate}

Evident by Fig.~\ref{fig:orbits}, encounter geometry also has a large impact on the perturbed orbit of the star. Encounters where the relative velocity has any azimuthal component will be the most effective at ejecting the star from resonance, as the change in $z$-component angular is momentum is what drives the change in slow action. Vertical encounters, with a non-zero vertical relative velocity component, are able to cause the star's vertical amplitude to change. This is not the case for purely parallel encounters with zero vertical relative velocity. Broadly speaking, perpendicular encounters primarily induce modest, symmetric deformations of the librating orbit. Even for $M\gtrsim10^{9}$M$_{\odot}$ subhaloes, the orbit remains trapped in co-rotation unless the mass is concentrated beyond the fiducial values given by Eq.~\ref{eq:scale_radius}. Parallel encounters are significantly more efficient at driving changes in the slow action, which is consistent with the analytic expectation that azimuthal velocity kicks dominate. In the more concentrated case shown in Fig.~\ref{fig:orbits}, the orbit transitions from libration to circulation, clearly crossing the separatrix.

By looking at the lower panels of Fig.~\ref{fig:orbits}, the change in $L_z$ is characterised by a large spike, and then some non-constant long term behaviour. In contrast, the change in $E_J$ is characterised by a jump to a new value that is sustained. This is expected, since $E_J$ is an integral of motion of the system. In the example shown for the parallel encounter (blue), there are two spikes in $L_z$ (or equivalently a second jump in $E_J$) as the subhalo crosses the path of the star twice. This is evident by the looping orbit in the co-rotation frame, shown in the top-right panels.

\begin{figure}
    \centering
    \includegraphics[width=\columnwidth]{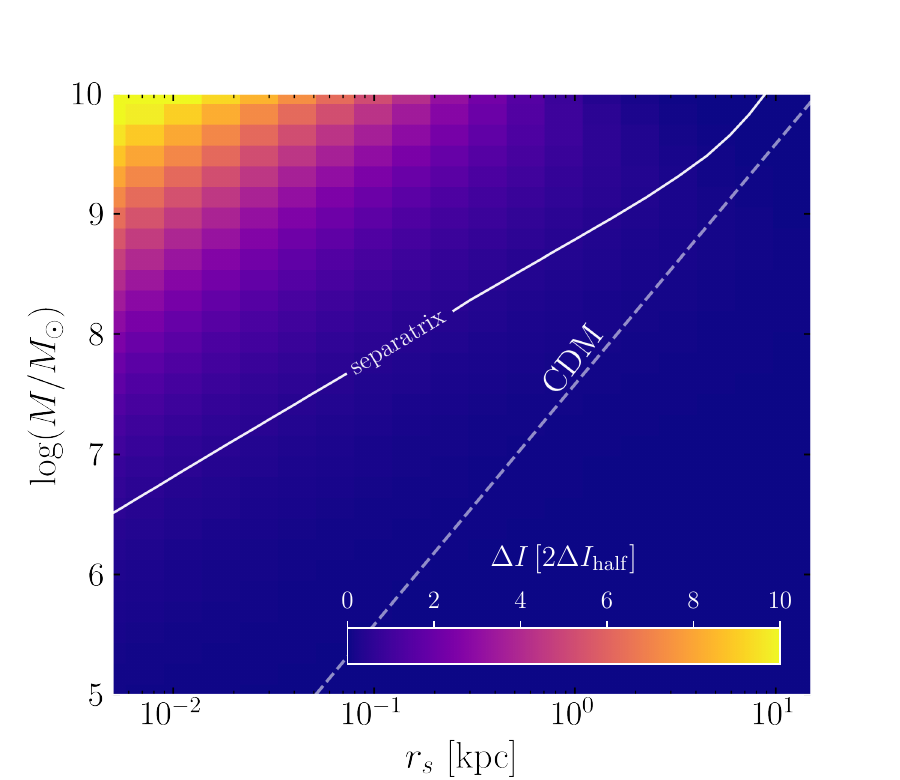}
    \caption{The change in slow action from a test particle simulation subhalo fly-by, evaluated at $t = t_{\rm enc} + 0.025$ Gyr for a vertical encounter, as a function of subhalo mass and scale radius. The dashed white line shows the $r_s-M$ relation given by Eq.~\ref{eq:scale_radius} and the solid white  line is a contour in $\Delta I$ which shows the separatrix width ($\Delta I = 2\Delta I_{\rm half}$). For each encounter, the impact parameter $b=r_s$, and the subhalo vertical velocity is $w_0 = 200$ km/s.}
    \label{fig:DeltaI_grid}
\end{figure}

Lastly, in Fig.~\ref{fig:DeltaI_grid} we present a summary of single subhalo encounter set-up by plotting the change in slow action for a range of scale radii and masses. As before, we evaluate the slow action at at $t = t_{\rm enc} + 0.025\,{\rm Gyr}$. This is not meant as a comprehensive overview, as there are many free parameters that can be tweaked. Instead this is meant to give an impression as to what regions of a subhalo's intrinsic property parameter space we expect a resonant star to experience a large perturbation. For each encounter we fix the geometry to a vertical fly-by with impact parameter $b = r_s$ and relative velocity $w_0 = 200$ km/s, such that the only free variables are the intrinsic properties of the perturbing subhalo. Unsurprisingly, the subhalo imparts the most change when it high mass and low scale radius (i.e. very concentrated). For a given mass, as the scale radius increases, the change in slow action decreases. At no point does equation the $r_s - M$ relationship given by Eq.~\ref{eq:scale_radius} (the white dashed line labelled as CDM) intersect with the separatrix (white solid line). For vertical encounters, using Eq.~\ref{eq:scale_radius}, we expect no Plummer sphere subhalo of mass $M<10^{10}$ M$_{\odot}$ to be capable of ejecting a resonant star. 

Therefore, the key implication of the results in this section is that only a restricted region of intrinsic parameter space permits single-encounter resonance destruction. Even when the impact parameter is chosen to maximise the impulse ($b = r_s$), and the geometry is favourable, the resulting change in slow action $\Delta I$ remains below the threshold required to move the resonant star beyond the separatrix. This motivates the consideration of a population of subhaloes instead of a single perturbing impact.

\section{Diffusion of resonances by many subhaloes}\label{sec:many_subhaloes}

In the previous section, we justified our usage of impulse approximation for a single subhalo impact albeit finding that a single subhalo has limited ability to eject a star from co-rotation resonance with the bar. In this section, we proceed to use the $\Delta I_{\rm sh} (M,b,u)$ in a diffusion model to calculate the impact of a \textit{population} of subhaloes, and define find the \textit{diffusion coefficient}, $D_{II}$, for this population. By choosing some subhalo mass function (SHMF) (and assuming sensible velocity distribution and impact parameter distribution) we may explore whether the Galactic subhalo population is capable of imparting enough change in action to eject a star from its resonance. While the previous section showed that only a very massive individual subhalo can cause a separatrix crossing. $\Delta I_{\rm sh} >\Delta I_{\rm half}$, we anticipate that the cumulative effect at the lower end of the SHMF may be sufficient to do so. By finding the cumulative impact of the SHMF (or certain mass ranges within it) we may draw conclusions from the existence of the co-rotation resonance.

For understanding the impact of many subhaloes on resonant stars, it is appropriate to use the Fokker-Planck approximation \citep[e.g. see §7.4 of][]{binneyandtremaine2008}. Namely, we can capture the impact of many low-mass subhaloes by their diffusion of the resonance stars distribution function $f(\bm{J'},t)$. In the Fokker-Planck approximation (in terms of actions), the full evolution of the distribution function is governed by
\begin{equation}
    \frac{\partial f}{\partial t} = -\sum_i \frac{\partial}{\partial J'_i}\Bigl\{ D_if\Bigl\} + \sum_{ij} \frac{\partial^2}{\partial J'_i \partial J'_j}\Bigl\{D_{ij} f\Bigl\},
\end{equation}
which is analogous to the diffusion equation with (orbit averaged) drift coefficients $D_i$ and (orbit averaged) diffusion coefficients $D_{ij}$, where $i,j$ run over the components of the action vector $\bm{J}'$, corresponding to the fast and slow actions. These coefficients describe the systematic drift and growth of variance of the actions, respectively. Critically for our context, escape from resonance depends only on whether the resonant star's slow action diffuses across the separatrix. Moreover, we consider the subhaloes as an isotropic distribution of perturbers. 

To consider escape from resonance, only diffusion in the slow action $I$ is relevant, since a star leaves the resonance once its slow action diffuses beyond the separatrix width $\Delta I_{\rm half}$. Rather than solving an explicit transport equation, we characterise the effect of the subhalo population entirely through the slow-action diffusion coefficient $D_{II}$:
\begin{equation}\label{eq:D_definition}
    D_{II} = \lim_{t\rightarrow0}\frac{\langle\Delta I_{\rm sh}^2\rangle}{2\Delta t} = \frac{1}{2}\int d\Gamma 
    \langle\Delta I_{\rm sh}^2\rangle,
\end{equation}
for a subhalo differential encounter rate $d\Gamma$, and mean square slow-action kick $\langle\Delta I_{\rm sh}^2\rangle$. The timescale for a particle in a bath of subhaloes to diffuse across the full resonance (i.e. twice the half width $\Delta I_{\rm half}$) is then given by,
\begin{equation}\label{eq:t_diff}
    t_{\rm diff} = \frac{(2\Delta I_{\rm half})^2}{2D_{II}} = (2\Delta I_{\rm half})^2 \left[\int d\Gamma \langle\Delta I_{\rm sh}^2\rangle\right]^{-1}.
\end{equation}
The diffusion timescale $t_{\rm diff}$ is sensitive to both the potential of the individual subhaloes (through $\Delta I_{\rm sh}$) and their population properties (through $d\Gamma$), allowing us to theoretically probe different dark matter models in few different ways. In the context of using resonant features as a probe of dark matter content of the Milky Way, we are interested in situations where the diffusion timescale is shorter than the lifetime of the resonant feature (or by proxy, the age of the Galactic bar $t_{\rm age}$). If $t_{\rm diff} < t_{\rm age}$, then the resonant feature should not exist under the assumption that no new stars are captured. Additionally, an understanding of the rate of change of the dispersion of the stars in the resonance would allow one to put constraints on the subhalo populations by comparison to the observed width of the resonance. However, this is beyond the scope of this work and likely not yet possible from a data perspective.
Using the diffusion coefficient we can also define the rms change in the slow action from a process with diffusion coefficient $D_{II}$ (after a total time $t$):
\begin{equation}\label{eq:total_I}
    \Delta I_{\rm rms}(t) = \sqrt{2 D_{II} t}.
\end{equation}
For a population of subhaloes with local spatial number density per unit mass $dn / dM$ (i.e. $dn/dM\times dM$ is the number density of subhaloes in the mass interval $M\rightarrow M+dM$) and a subhalo velocity distribution $f(\bm{w})$, the differential encounter rate (in the resonant star's reference frame) can be written as
\begin{equation}
    d\Gamma = \frac{dn}{dM} \, dM \, f(\bm{w}) d^3\bm{w} \, u \, (2\pi b\, db),
    \label{eq:dGamma}
\end{equation}
where $2\pi b\,db$ represents the ring-shaped area element in impact-parameter space, and $u$ is the aforementioned relative speed of the star and subhaloes. 

\subsection{Subhalo velocity distribution}

Here we could choose the velocity distribution of dark matter subhaloes in the inertial frame to be an isotropic Maxwellian distribution with velocity dispersion $\sigma$:
\begin{equation}\label{eq:f_velocity}
    f(\bm{w}) = \left(\frac{1}{2\pi \sigma^2}\right)^{3/2} \exp{\left( -\frac{w^2}{2\sigma^2}\right)}.
\end{equation} 
Since this is the distribution of velocities in the inertial frame, we need to account for the shift from the velocity of the resonance star $\bm{v} = (0,v_*,0)$. Because $\bm{w} = \bm{u} - \bm{v}$, the change of integration variables is simply $d^3\bm{w} \rightarrow d^3\bm{u}$. Rewriting the distribution in terms of the relative velocity, and integrating over the angles, gives us the speed distribution (see Appendix~\ref{appendix:velocity} for details),
\begin{equation}\label{eq:f(u)}
    f(u) = \frac{u}{\sigma v_*\sqrt{2\pi}} \left[ \exp \left(-\frac{(u - v_*)^2}{2\sigma^2}\right) - \exp \left(-\frac{(u + v_*)^2}{2\sigma^2}\right) \right].
\end{equation}

\subsection{Impact parameter distribution}

As a reminder, the impact parameter vector $\bm{b}$ is defined as the separation between the star and the subhalo at the time of closest approach. By construction, $\bm{b}\cdot\bm{u}=0$, which confines $\bm{b}$ to the 2-d plane perpendicular to relative velocity of the star and subhalo $\bm{u}$. We subsequently refer to this plane as the \textit{encounter plane}.

In an isotropic 3-dimensional distribution, each component of $\bm{b}$ has equal variance, enabling one to take the average $\langle b_{\varphi}^2\rangle = b^2/3$. However, since that is not true here, this factor needs to be corrected. More precisely, for a fixed relative velocity $\bm{u}$, the impact parameters are isotropically distributed within the encounter plane. As a result, the distribution of $\bm{b}$ is 2-dimensional, and any component of $\bm{b}$ must be understood as a projection within the encounter plane, conditional on $\bm{u}$. 

Therefore, we first condition on $\bm{u}$ and then average over the 2-dimensional isotropic distribution. This yields an expression for the conditional expectation value $\langle b_{\varphi}^2 | \bm{u}\rangle$. This expression captures the geometric coupling between the azimuthal direction and the 2-dimensional encounter plane.

This procedure absorbs all geometric dependence on the encounter orientation into a relative velocity-dependent factor, allowing the integral of $d\Gamma$ over impact parameter and relative velocity to be separated, despite a dependence of $\bm{b}$ on the subhalo velocity $\bm{w}$.

We previously found the impact parameter in the azimuthal direction by simply taking $b_{\varphi} = \bm{b}\cdot\bm{e}_{\varphi}$, where $\bm{e}_{\varphi}$ the azimuthal direction relevant for the kick. To account for the movement of the star, we decompose the unit vector into the parts parallel and perpendicular to the star's relative velocity unit vector $\bm{\hat{u}}$:
\begin{equation}
    \bm{e}_{\varphi} = (\bm{e}_{\varphi}\cdot\bm{\hat{u}})\bm{\hat{u}} + \bm{e}_{\varphi,\perp}
\end{equation}
where $\bm{e}_{\varphi,\perp} \cdot \bm{\hat{u}} = 0$, and $|\bm{e}_{\varphi, \perp}|^2 = 1 - (\bm{e}_{\varphi} \cdot \bm{\hat{u}})^2$. Altogether, with the 2-dimensional encounter plane isotropy, the azimuthal impact parameter conditioned on the star's relative velocity $\bm{u}$ is:
\begin{equation}
    \langle b_{\varphi}^2 | \bm{u}\rangle = \frac{b^2}{2} \left(1 - (\bm{e}_{\varphi} \cdot \bm{\hat{u}})^2\right).
\end{equation}
We subsequently average over all possible relative velocity directions $\bm{u}$, under some assumption about the distribution of $\bm{\hat{u}}$. A good way to describe a distribution of vectors with a preferred direction is the Fisher distribution \citep[][]{fisher1953dispersion}:
\begin{equation}
    f_{\bm{\hat{u}}}(\bm{\hat{u}}) = \frac{\xi}{4\pi\sinh\xi}\exp\left(\xi\,\bm{e}_{\varphi}\cdot\bm{\hat{u}}\right),
\end{equation}
where $\xi = uv_*/\sigma^2$. Integrating over this distribution ultimately results in a relative velocity dependent impact parameter (see Appendix~\ref{appendix:impact} for details):
\begin{equation}\label{eq:b_correction}
    \langle b_{\varphi}^2 | \bm{u}\rangle =b^2 \left[\frac{\coth(\xi)}{\xi} - \frac{1}{\xi^2}\right]
\end{equation}
We can write this in a more compact form:
\begin{equation}
    \langle b_{\varphi}^2 | \bm{u}\rangle = b^2 \lambda(u)
\end{equation}
where the correction factor $\lambda$ is only a function of the relative velocity $u$, since we fix the subhalo dispersion $\sigma$ and the star's velocity $v_* = v_0$. In the limit of $v_*\rightarrow0$, $\lambda(u)\rightarrow1/3$ and so we recover the original 3-dimensional isotropic assumption.

\subsection{Subhalo number distribution}

The local subhalo number density $dn / dM$ describes the number of subhaloes per unit volume and per unit mass interval at the location of interest (e.g. near co-rotation). It is related to the subhalo mass function $dN / dM$—which gives the total number of subhaloes per unit mass within the host halo—through the normalised radial distribution of subhaloes $P(r)$:
\begin{equation}
    \frac{dn(r)}{dM} = \frac{dN}{dM} \, P(r),
    \label{eq:dn_dM_relation}
\end{equation}
where $P(r)$ is given by \citep[][]{springel2008aquarius}
\begin{equation}
    P(r) \propto \exp \left(-\frac{2}{\alpha}\left(\left(\frac{r}{r_{-2}}\right)^{\alpha} - 1\right)\right),
\end{equation}
normalized such that $\smallint P(r) \, 4\pi r^2 \, dr = 1$, where $\alpha = 0.678$ and $r_{-2} = 0.81 r_{200}$. 

\subsection{Diffusion coefficient}

To compute the diffusion coefficient, we need the differential encounter rate $d\Gamma$ and the mean square slow-action kick $\langle(\Delta I)^2\rangle$. For practical purposes the differential encounter rate is written as
\begin{equation}
    d\Gamma = P(r_{\rm CR})\frac{dN}{dM}dM \, f(u) du\, u \, (2\pi b\,db),
\end{equation}
and the mean squared slow-action kick is
\begin{equation}
     \langle(\Delta I_{\rm sh})^2 \rangle= \left(\frac{r}{n_{\varphi}}\right)^2 \left(\frac{2GM}{u}\right)^2 \frac{\langle b_{\varphi}^2 | \bm{u}\rangle}{(b^2 + r_{\rm s}^2)^2}.
\end{equation}
In the integral over the impact parameter, we choose $b_{\rm max}$ to be the virial radius of the Milky Way, $b_{\rm max} =r_{\rm 200} = 210$ kpc. To perform the integral in Eq.~\ref{eq:D_definition}, we separate the integral over impact parameter $b$ and relative velocity $u$ into parameters $\mathcal{B}$ and $\mathcal{V}$ respectively. The integral over the impact parameter evaluates to,
\begin{align}
    \mathcal{B}(M) =
    \ln\left(
    \frac{b_{\max}^{2}+r_{\rm s}^{2}}{r_{\rm s}^2}
    \right)
    +
    \frac{r_{\rm s}^2}{b_{\max}^{2}+r_{\rm s}^{2}}
    - 1
\end{align}
This can be simplified since $b_{\max}^2 \gg r_s^2$ for all values of $M$:
\begin{equation}
    \mathcal{B}(M) \approx \ln\left(\frac{b_{\max}^2}{r_s^2}\right) -1.
\end{equation}
Using Eq.~\ref{eq:scale_radius}, and defining $\kappa = 1.62^2 \times 10^{-8}$, we can express $\mathcal{B}$ as:
\begin{equation}
    \mathcal{B}(M) = \mathcal{B}_{\rm P} -\ln(M), 
\end{equation}
where $\mathcal{B}_{\rm P} = \ln(b_{\max}^2 / \kappa) - 1 = 27.15$. We have denoted this constant with a subscript P to indicate that its value is specific to the Plummer sphere potential. This value changes if we assume a new potential (and therefore a new $r_s - M$ relation). We discuss the case of a Hernquist potential in the Appendix.

Separating out the parts of the differential interaction rate that rely only on the velocity, we find that the parameter $\mathcal{V}$ is given by,
\begin{equation}
    \mathcal{V}(v_*;\sigma)  = \int^{\infty}_{0} \lambda(u)f(u) u^{-1} du.
\end{equation}
The integral without the $\lambda(u)$ term has a well established result \citep[][]{chandrasekhar1943dynamical, rosenbluth1957fokker, spitzer1987}:
\begin{equation}
    \int^{\infty}_{0} f(u) u^{-1} du = \frac{1}{v_*}\erf\left(\frac{v_*}{\sqrt{2}\sigma}\right),
\end{equation}
where $\erf$ is the error function. In the limit of $v_* \ll \sigma$ this reduces to $\sqrt{2/(\pi\sigma^2)}$, while in the limit of $v_* \gg \sigma$ this becomes $1/v_*$. However, because of the factor $\lambda(u)$, we will need to integrate $\mathcal{V}$ numerically.

Plugging in all the relevant expressions to Eq.~\ref{eq:D_definition} we get, 
\begin{equation}\label{eq:diffusion}
    D_{II} = 2\pi\left(\frac{r_{\rm CR}G}{n_{\rm \varphi}}\right)^2P(r_{\rm CR}) \times \mathcal{V}(v_*; \sigma) \times \mathcal{I}(M),
\end{equation}
where the integral $\mathcal{I}(M)$ is given by 
\begin{equation}\label{eq:mass_integral}
        \mathcal{I}(M) =\int_{M_{\rm min}}^{M} \frac{dN}{dM'}M'\,^2\mathcal{B}(M')\,dM',
\end{equation}
where the minimum mass in the integral is set to $M_{\rm min} = 10^{5}$ M$_{\odot}$ to be within a range for which the SHMF adopted from \citet[][]{springel2008aquarius} is valid. The maximum mass does not exceed $M=10^{10}$ M$_{\odot}$ for the same reason.
The integrals $\mathcal{I}$ and $\mathcal{V}$ are solved using the scipy \texttt{quad} function.

\subsection{Dark matter model}

\begin{figure}
    \centering
    \includegraphics[width=0.95\columnwidth]{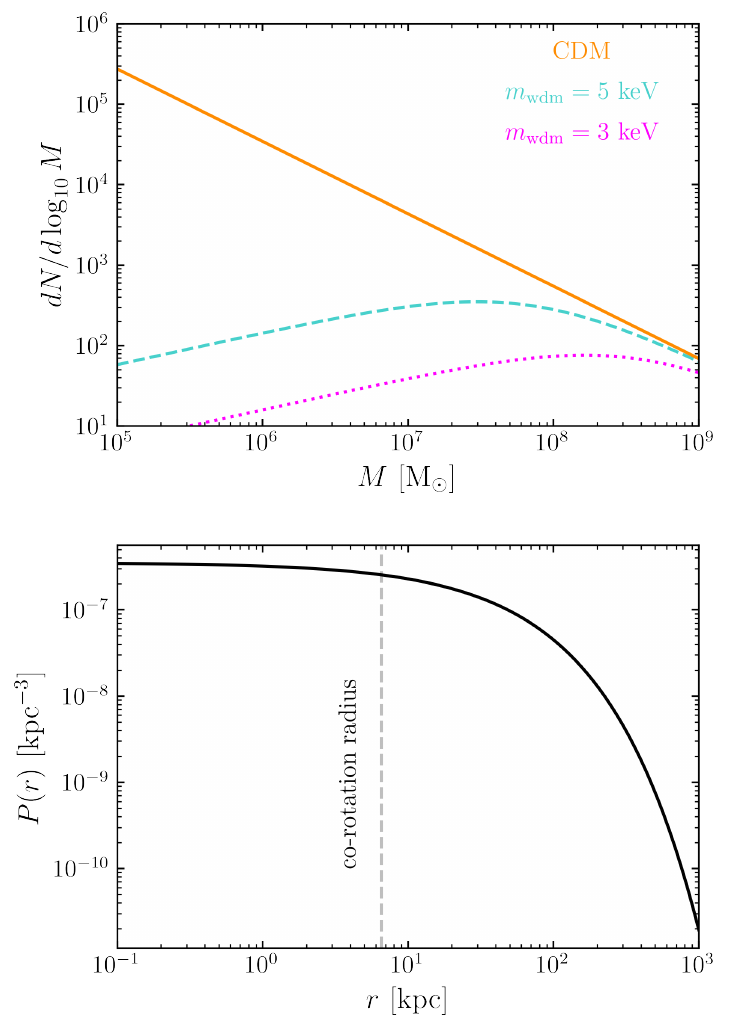}
    \caption{Subhalo population properties used in the analytical diffusion model. The top panel shows the subhalo mass function (SHMF) $dN/dM$ for three different dark matter models. The orange line shows the SHMF for CDM, the cyan and magenta line shows CDM-suppressed models which approximate a WDM model. The suppressed models simply reduce the amount of subhaloes at lower massses. The bottom panel shows the radial distribution of subhaloes $P(r)$, and the position of the co-rotation radius. Both the SHMF and $P(r)$ are taken from simulation Aq-A-1 in Aquarius suite \citep[][]{springel2008aquarius}.}
    \label{fig:dm_properties}
\end{figure}

The sensitivity to different dark matter models appears only through the diffusion coefficient $D_{II}$ (or equivalently the diffusion timescale $t_{\rm diff}$). Specifically, different models of dark matter will produce different forms of the SHMF, $dN/dM$. Alternatively, models such as self-interacting dark matter, which have little impact on the SHMF, present their impact through the relationship between the scale radius and mass (contained in $\mathcal{B}$). From cosmological simulations \citep[][]{springel2008aquarius}, the SHMF has been shown to take the form
\begin{equation}\label{eq:shmf}
    \frac{dN}{dM} = a_0 \left( \frac{M}{m_0} \right)^{-\alpha_{\rm DM}} \times \mathcal{S}(M),
\end{equation}
with some mass-dependent suppression factor $S(M)$ for models varying from CDM. Throughout this work we take the values of the parameters in Eq.~\ref{eq:shmf} to be $a_0 = 3.26\times10^{-5}$ M$_{\odot}^{-1}$, $m_0 = 2.25 \times 10^7$ M$_{\odot}$, and $\alpha_{\rm DM}=1.9$. We choose to take a typical form \citep[e.g.][]{lovell2014properties}:
\begin{equation}
    \mathcal{S}(M) = \left(1 + \frac{M_{\rm hm}}{M}\right)^{-1.3},
\end{equation}
where the transition mass $M_{\rm hm}$ sets the subhalo number supression relative to CDM, and is related to the WDM particle mass by \citep[][]{schneider2012nonlinear, viel2013warm},
\begin{equation}
    M_{\rm hm} =10^{10} h^{-1} \left( \frac{m_{\rm WDM}}{1 \,\rm keV}\right)^{-10/3} {\rm M}_{\odot},
\end{equation}
with $m_{\rm WDM}$ being the DM particle mass and $h=0.7$. Introducing this parameter modifies the integral $\mathcal{I}(M)$ to become:
\begin{equation}
        \mathcal{I}(M) =\int_{M_{\rm min}}^{M} \frac{dN}{dM'}M'\,^2\mathcal{B}(M')\,\mathcal{S}(M')\,dM'.
\end{equation}
In the top panel of Fig.~\ref{fig:dm_properties} we show the corresponding $dN/d\log M$ for the subhalo mass functions described by Eq.~\ref{eq:shmf} for a few values of $m_{\rm WDM}$. The orange line shows the CDM SHMF, with parameters chosen from the Aquarius simulations. The dashed cyan line and the dotted magenta line show the SHMF modified with a WDM particle mass of 5 keV and 3 keV respectively, to the emulate suppression resulting from a warm dark matter particle. In the bottom panel of Fig.~\ref{fig:dm_properties} we show the radial distribution of subhaloes $P(r)$, whose value at $r=r_{\rm CR}$ is used to calculate the subhalo number density at the co-rotation resonance for calculation of the diffusion coefficient. 

\section{Results \& Discussion}\label{sec:discussion}

In this section we present the implications of the subhalo diffusion coefficient $D_{II}$ (given in Eq.~\ref{eq:diffusion}). We express our results in terms of a few different parameters: (a) the diffusion coefficient itself, (b) a dimensionless diffusion strength $\Delta$, (c) the timescale of diffusion $t_{\rm diff}$, and (d) the total change in slow action $\Delta I$. Additionally, we show how the consequences of the subhalo diffusion change according to varying bar properties and dark matter models.

By the assumption of an isothermal, isotropic spherical halo with an flat rotation curve, we can relate the velocity dispersion to the circular velocity $\sigma \sim v_0/\sqrt{2}$ using Jeans equations. By taking $v_0 = 230$ km/s \citep[][]{eilers2019circular, ou2024dark}, it would be reasonable to assume $\sigma \sim 160$ km/s. However, knowing the precise value of the velocity dispersion at the local volume is critical to our method, as it directly scales the diffusion coefficient. Therefore, to place an upper limit on $D_{II}$, we assume a more conservative value of $\sigma = 200$ km/s. We take the maximum impact parameter to be the virial radius of the Milky Way, $b_{\rm max}= 210$ kpc \citep[e.g.][]{shen2022mass}. Wherever bar parameters are not explicitly stated, they can be assumed to be roughly the same as those in \citet[][]{chiba2021resonance}: (a) pattern speed of $\Omega_{\rm p} = 35$ km s$^{-1}$ kpc$^{-1}$, (b) a strength of $A=0.02$, and (c) a scale length of $r_{\rm b} = 1.6$ kpc.

\subsection{Diffusion coefficient}

Here we present the behaviour of the diffusion coefficient, and how it varies as we change the dark matter models. We show both the actual diffusion coefficient $D_{II}$, and also the dimensionless diffusion strength $\Delta$ (introduced by \citet[][]{hamilton2023galactic}).

\begin{figure}
    \centering
    \includegraphics[width=0.95\columnwidth]{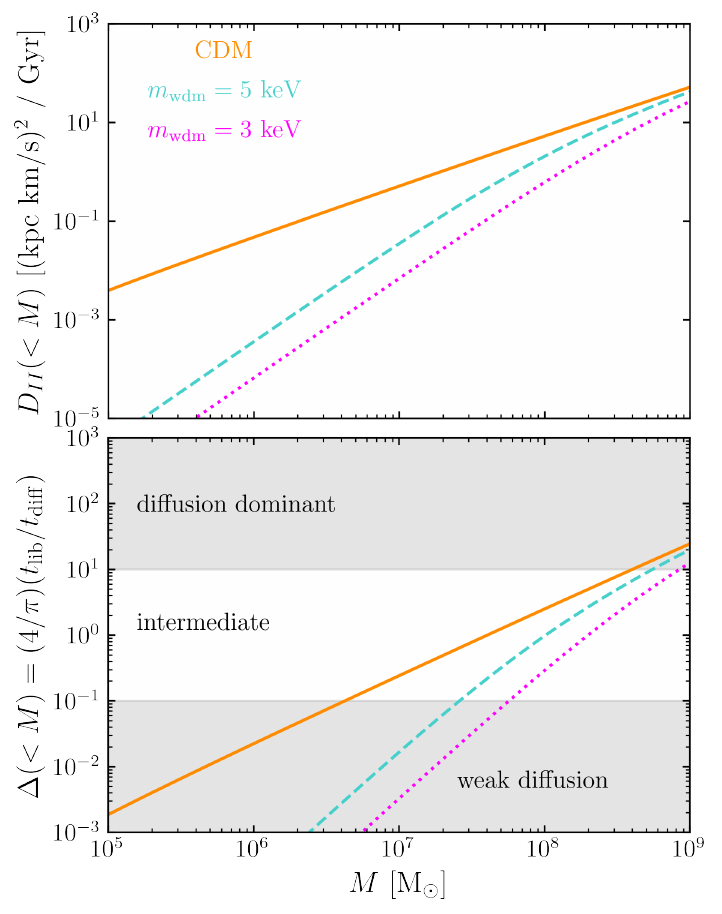}
    \caption{diffusion coefficient $D_{II}$ (top panel) and dimensionless diffusion strength $\Delta$ (bottom panel) integrated over mass values up to $M$. The subhalo population parameters are chosen to match the three types of $dN/d\log M$ and $P(r)$ shown in Fig.~\ref{fig:dm_properties}.}
    \label{fig:diffusion_coefficient}
\end{figure}

In Fig.~\ref{fig:diffusion_coefficient}, we illustrate the result of integrating the diffusion coefficient $D_{II}$ up to different values of mass $M$, to obtain the cumulative diffusion coefficient for all subhaloes up to and below mass that mass $M$. In the top panel of Fig.~\ref{fig:diffusion_coefficient}, we show this for three different dark matter models. Specifically, we compute $D_{II}$ for CDM, and two ``warm'' dark matter models, whose SHMF is given by Eq.~\ref{eq:shmf} with particle masses of $m_{\rm WDM}=3$ keV and  $m_{\rm WDM} = 5$ keV. These values are chosen as the bound the current observational constraints as described in the introduction. As expected, the warm dark matter models have a suppressed $D_{II}$ compare to CDM at lower masses, according to how their respectively SHMFs are suppressed.

Since the meaning behind the values of the diffusion coefficient $D_{II}$ are difficult to interpret physically, we make use of the dimensionless diffusion strength \citep[][]{hamilton2023galactic},
\begin{equation}\label{eq:dimensionless_diffusion_strength}
    \Delta = \frac{4}{\pi}\frac{t_{\rm lib}}{t_{\rm diff}}.
\end{equation}
$\Delta$ measures how many libration periods elapse before diffusion moves a star across the separatrix. When $\Delta \ll 1$ the diffusion is very weak and the resonant behaviour dominates. Conversely, when $\Delta \gg 1$ the diffusion process dominates and the imprint of any resonance is washed away. For intermediate regimes, where $\Delta \sim 0.1 - 10$ the diffusive process is already strong enough to produce noticeable deviations from purely collisionless behaviour. By examination of the bottom panel of Fig.~\ref{fig:diffusion_coefficient}, we could expect to see an imprint from the subhaloes on the resonant features below $M = 10^8$ M$_{\odot}$. Beyond $M \gtrsim 10^8$ M$_{\odot}$, we certainly expect the diffusive process to dominate. However, once we approach masses above $M=10^9$ M$_{\odot}$, where the total number of subhaloes in the Galaxy is expected to be $N<100$, we can probably not rely on the assumption of a diffusive subhalo bath regime to hold.

\subsection{Timescale of diffusion}

Since the value of the diffusion coefficient itself is difficult to place into a Galactic context, here we express it through different means which are themselves detailed in Sec.~\ref{sec:many_subhaloes}. 

The most intuitive way to understand the impact of the Galactic subhalo population on the stars in resonance with the bar is through the timescale of diffusion. If a population of subhaloes has a diffusion timescale $t_{\rm diff}$ that is shorter than the lifetime of the bar resonance itself, then it is plausible that the resonance should not be able to exist. This obviously requires one to know the lifetime of the bar resonance. We use an estimate for the age of the bar itself as a proxy. 

The diffusion timescale is given in Eq.~\ref{eq:t_diff}, where we can see that it scales with the inverse of the diffusion coefficient and depends on the bar parameters through the resonance half-width $\Delta I_{\rm half}$.

\begin{figure}
    \centering
    \includegraphics[width=0.95\columnwidth]{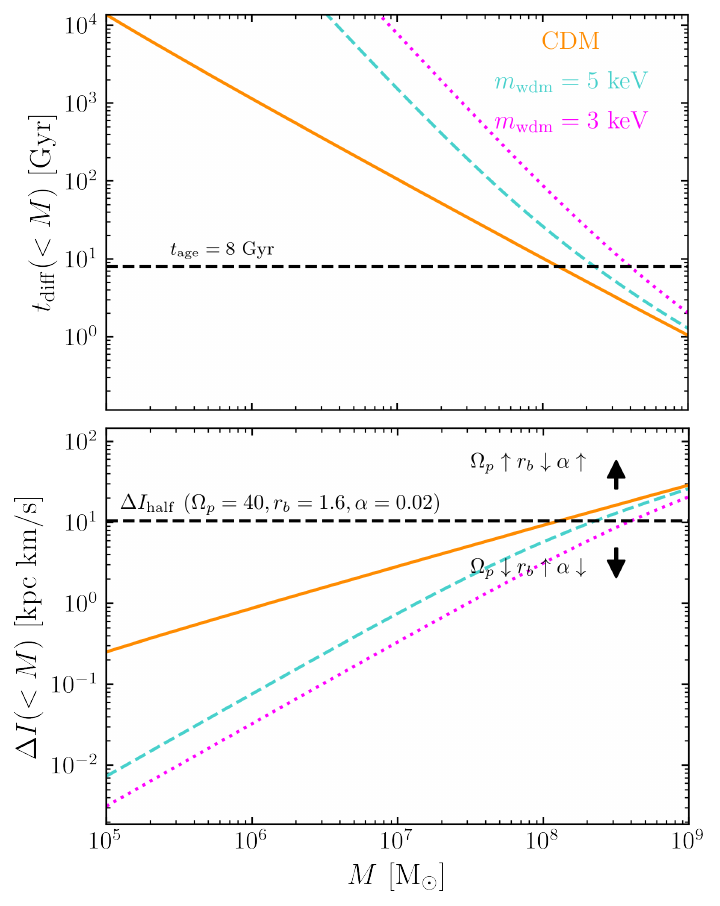}
    \caption{Diffusion timescale (top panel) and total change in slow action (bottom panel). The diffusion timescale is calculated assuming bar paramters described earlier, and diffusion coefficient shown in Fig.~\ref{fig:diffusion_coefficient}. The black dashed line in the top panel shows an estimate for the bar age at $t_{\rm age} = 8$ Gyr, which is a proxy for the timescale required to diffuse away the bar resonance. In the bottom panel, the total change in slow action assumes $8$ Gyr of diffusion. The black horizontal lines show the half-width for a variety of bar parameters, and indicate the required value of $\Delta I$ to diffuse away the bar resonance.}
    \label{fig:timescale}
\end{figure}

In the top panel of Fig.~\ref{fig:timescale}, we present the cumulative diffusion timescale (integrated from $M_{\rm min}~=~10^5$~M$_{\odot}$ up to $M_{\rm max}~=~M$) for three different dark matter models. The dashed horizontal line represents the age of the bar $t_{\rm age}$ \citep[e.g.][]{sanders2024epoch} (used as a proxy for the lifetime of the resonance). The range of subhalo masses where $t_{\rm diff}(<M) < t_{\rm age}$ is the region for which the subhaloes would have diffused away the parent orbit of the resonance. The orange line therefore shows two key details for CDM. Firstly, the resonant feature is insensitive to the population of subhaloes up to $M\sim10^7$M$_{\odot}$. Secondly, assuming the diffusive region applies, the population of subhaloes above $M\gtrsim10^8$M$_{\odot}$ could have plausibly diffused the resonance away. Given that the co-rotation resonance has possibly been detected in the stellar halo (and more generally known to exist in the disk), this would imply some level of suppression from CDM at the location of the co-rotation resonance. For the other dark matter models, we see an increase in the diffusion timescale for lower mass subhaloes. This is to be expected, as these models simply reduce the number of subhaloes at these lower masses, which would reduce the diffusion coefficient. 

\begin{figure}
    \centering
    \includegraphics[width=0.95\columnwidth]{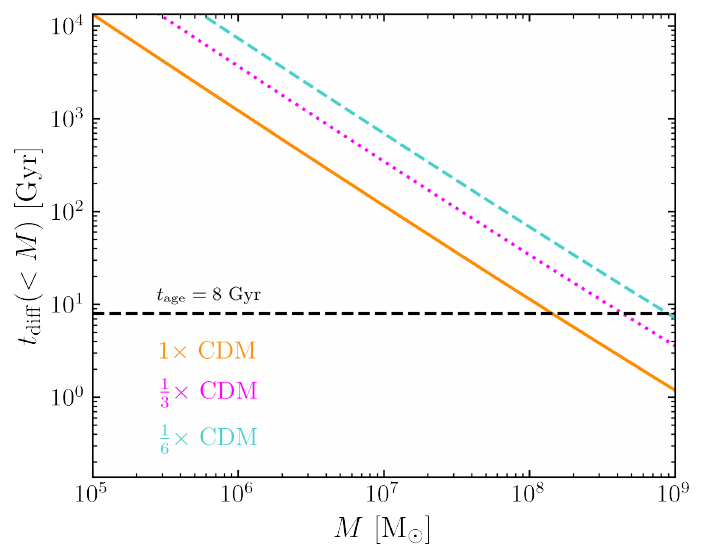}
    \caption{The re-scaled diffusion timescale, after changing the density of subhaloes at co-rotation $P(r=r_{\rm CR})$ by different factors. The solid orange line is normal CDM, the dotted magenta line shows CDM scaled down to $1/3$ the normal density, and the dashed turquoise line shows the CDM scaled down to $1/6$ the normal density. These scalings are chosen so that the diffusion timescale does not become shorter than the bar's age for all masses up to $10^8$ M$_{\odot}$ and $10^9$ M$_{\odot}$ respectively. The existence of a bar resonance (assuming an age of $8$ Gyr) should imply some suppression to CDM expectations.}
    \label{fig:constraints}
\end{figure}

To understand the level of suppression from CDM predictions, we re-plot the diffusion timescale but with various levels of modification to the subhalo density at the co-rotation resonance $P(r=r_{\rm CR})$. In Fig.~\ref{fig:constraints} we present how the diffusion timescale is shifted when the subhalo density is scaled to be $1/3$ of its CDM value, and $1/6$ of the its CDM value. These modifications cause the diffusion timescale to be closer to the age of the bar for all subhaloes up to $M=10^8$ M$_{\odot}$ and $M=10^9$ M$_{\odot}$ respectively. With it being more likely that the diffusive subhalo bath regime assumption breaks at higher masses, we consider it more reasonable that the existence a stellar population trapped at the co-rotation resonance implies a suppression at the $r=r_{\rm CR}$ up to around $1/6$ of the expected CDM value. This corroborates with the suppression factor of roughly $1/3$ due to tidal disruption by the disk at the solar radii \citep[][]{d'onghia2009substructure, erkal2016number}. The existence a stellar population trapped at the co-rotation resonance therefore provides an independent constraint on the subhalo suppression near the solar volume that is comparable to other works, and can be used as a calibration factor to such suppression.

\subsection{Impact of bar properties}

As an additional way to understand the impact of the Galactic subhalo bath, we plot the cumulative total change in slow action $\Delta I$ (for all the subhaloes up to mass $M$). Instead of thinking in timescales, we can consider by how much the slow action of a resonant star has changed after a certain time (we take $t=8$ Gyr). This total change can then be compared to the half-width of the resonance, and if this half width is exceeded, the resonance can be diffused away. The total change in slow action after some time $t$, under diffusion coefficient $D_{II}$, is presented by equation Eq.~\ref{eq:total_I}. This re-formatting of the diffusion timescale makes it easier to understand how changing the bar properties impacts our result (since $\Delta I$ is not depending on bar properties).

In the bottom panel of Fig.~\ref{fig:timescale}, we plot $\Delta I (<M)$ for the same three dark matter models, and compare this with three different configurations for the bar properties. This plot is intended to present how the half width varies as we tweak the bar properties. The large black arrows show that: (a) increasing the pattern speed increases the half-width, (b) increase the bar length decreases the half-width, and (c) increasing the bar strength increases the half-width. As this plot is just a re-formulation of the diffusion timescale result, we see that the middle dashed line shows again that the resonance is insensitive to subhaloes up to $M\lesssim10^7$ M$_{\odot}$. Changing the bar properties has interesting implications. A faster rotating, longer, or weaker strength bar makes the resonances more sensitive to lower mass subhaloes, and thus makes the co-rotation resonance a better subhalo detector. The likely evolution (i.e. slowing down) of the bar makes this much more complicated, and we will explore it in future work. 

\section{Conclusions}\label{sec:summary}

In this section we first summarise the work presented in this paper and then discuss future work that will develop this framework further.

\begin{figure}
\centering

\begin{minipage}[t]{0.45\textwidth}
\centering
\begin{tikzpicture}[x=1cm,y=1cm,baseline=(bb.south)]
  \path[use as bounding box] (0,0) rectangle (7,6);
  \coordinate (bb) at (0,0);

  \draw[->] (0,0) -- (0,6);
  \draw[->] (0,0) -- (7,0);
  \node[black, rotate=90, font=\large] at (-0.4, 3.5) {Angular Momentum};

  \fill[gray!30] (2,0.75) rectangle (6,1.25);  
  \fill[gray!30] (2,2.80) rectangle (6,3.20);  
  \fill[gray!30] (2,3.85) rectangle (6,4.15);  
  \fill[gray!30] (2,4.40) rectangle (6,4.60);  
  \fill[gray!30] (2,4.68) rectangle (6,4.82);  

  \node[anchor=west] at (6,1.00) {CR};
  \node[anchor=west] at (6,3.00) {OLR};
  \node[anchor=west] at (6,4.00) {$n_r\geq2$};

  \draw[dashed] (0,5) -- (7,5);

  \draw[->, red, ultra thick] (3,0.85) -- (3,3.0);
  \draw[->, red, ultra thick] (3.8,0.85) -- (3.8,1.2);
  \draw[->, red, ultra thick] (4.6,0.85) -- (4.6,2.0);
  \draw[->, red, ultra thick] (5.4,0.90) -- (5.4,0.4);
  \draw[<->, black, thin] (1.85,0.75) -- (1.85,1.25);

  \node[red,left] at (2.8,2.0) {Subhalo impact, $\Delta I_{\rm sh}$};
  \node[black,left] at (1.8,1.0) {$2\,\Delta I_{\rm half}$};
\end{tikzpicture}
\end{minipage}

\caption{Schematic diagram for the impact of a subhalo fly-by on a star in resonance with the Galactic bar. A subhalo may cause a star in co-rotation resonance to escape beyond the separatrix, get trapped by another resonance, or remain in co-rotation. The bath of subhaloes may deplete the co-rotation resonance (CR) and populate higher-order resonances. Higher order resonances have narrower half widths, and so are diffused away easier by the same subhalo perturbation. However, this is somewhat counteracted by the reduced subhalo number density at larger Galactic radii.}
\label{fig:atom}
\end{figure}
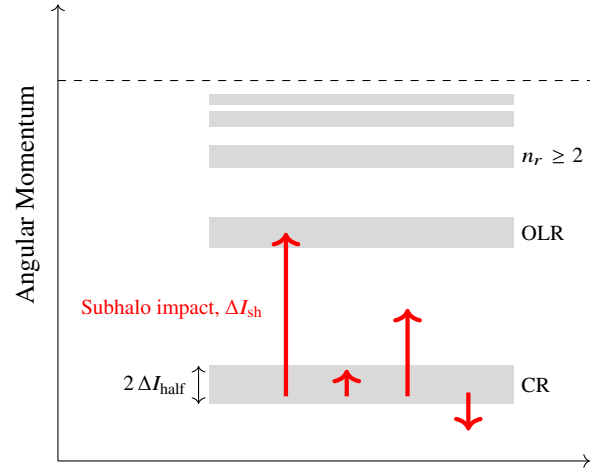

\subsection{Summary}

We introduce and discuss the usage of bar-induced resonances in the Galaxy as a tool for detecting low-mass dark matter subhaloes with the ultimate aim of constraining the subhalo mass function (SMHF). As a first proof-of-concept, we take an analytical approach to estimate the impact of the Galactic subhalo population on bar-induced resonant features in the Milky Way. Namely, we treat every individual subhalo encounter as an impulse, and aggregate their behaviour as a diffusive process. Subsequently, we obtain an expression for the diffusion coefficient $D_{II}$, the diffusion timescale $t_{\rm diff}$, and other related parameters. The problem simplifies down to a questions of whether a certain population of subhaloes changes the slow action $I$ of a resonant star enough to exceed the half-width $\Delta I_{\rm half}$ of the resonance. Consideration of timescales makes the question more intuitive. Namely, it enables us to formulate the problem as a comparison between the timescale of diffusion and the lifetime of the bar resonances, in which the simple existence or non-existence of resonant features in the Galaxy provide constraints on the subhalo population. The analytical work is complemented by basic test particle simulations that act both as a visual aid, and are used to understand the limitations of our analytical model.

Using the subhalo diffusion coefficient, we find that for CDM the co-rotation resonance is likely insensitive to subhaloes below $M~<~10^7$~M$_{\odot}$, slightly heated by the cumulative subhaloes up to $M~\sim~10^8$~M$_{\odot}$, and likely dissipated when accounting for subhaloes with $M~\gtrsim~10^8$~M$_{\odot}$. By instead comparing the timescales of diffusion with the age of the bar (as a proxy for the lifetime of the resonance), we similarly predict that the CDM predicted subhalo mass function should have dissipated the co-rotation resonance away. This is remedied if we assume that the density of subhaloes at the co-rotation resonance is 1/6 that predicted by CDM \citep[][]{springel2008aquarius}. This suppression to the local density corroborates other studies that assume a suppression of this amount at the local volume as a result of tidal stripping \citep[e.g.][]{d'onghia2009substructure, erkal2016number}.

With the inclusion of a WDM suppression term to the SHMF, there is a reduced diffusion coefficient relative to the CDM, as fewer low-mass subhaloes contribute to the cumulative perturbation of resonant stellar orbits. Consequently, the efficiency with which stars are scattered across the resonant separatrix is diminished in WDM models, particularly below $M \sim 10^8\,{\rm M_\odot}$ where the suppression becomes significant.

Varying the bar properties impacts the sensitivity of the co-rotation resonance to subhalo perturbations through changes in the width of the resonance in slow action space. Specifically, (a) increasing the pattern speed increases the resonance half-width, (b) increasing the bar length decreases the resonance width, and (c) increasing the bar strength increases the width. 

We summarise the framework presented in this work by comparison to the atomic model, where resonances are analogous to energy levels (with some intrinsic width), and subhaloes are equivalent to photons which can induce transitions. Fig.~\ref{fig:atom} presents this comparison as a schematic. The exact analogy breaks down in a few ways (notably that subhalo impacts are not quantized), but we believe this paints a comprehensible picture that nicely summarises the idea.

\subsection{Limitations}

While the framework developed in this work provides a first proof-of-concept for using bar resonances as probes of the Galactic subhalo population, several simplifying assumptions limit the scope and precision of our results.

For analytic tractability we restrict our attention to stars on circular planar orbits and focus exclusively on the co-rotation resonance. In reality, halo stars (the stars most relevant for this framework) exhibit a wide range of eccentricities and vertical excursions. The slow action for non-planar orbits generally mixes $J_r$, $L_z$, and $J_z$. Consequently, the mapping $\Delta I = \Delta L_z / n_\varphi$ adopted in our work is exact only for the simplified configuration studied. Extending the formalism to fully three-dimensional phase space will likely modify the quantitative diffusion coefficients and resonance widths.

All subhaloes are represented using simplified Plummer sphere profiles with a $r_s - M$ relationship motivated by cosmological simulations. Real subhaloes exhibit scatter in concentration, tidal truncation, and possible baryonic modification. Moreover, the local subhalo number density is uncertain due to tidal disruption by the Galactic disc. Our results therefore depend on the adopted structural relations and radial distribution.

We treat the subhalo population as a homogeneous stochastic bath. This assumes that many weak encounters dominate over rare strong ones. While this is appropriate for low-mass subhaloes, it becomes questionable at the high-mass end ($M \gtrsim 10^9\,M_\odot$), where the expected number of objects is small. In this regime, the dynamics may be dominated by discrete, rare events rather than continuous diffusion, and a purely diffusive description may be inadequate.

Lastly, the bar is treated as a rigid, steadily rotating perturbation with fixed pattern speed, strength, and scale length. However, the Galactic bar has likely evolved over time. This implications for resonant features since bar slow-down shifts resonance locations over time and may alter resonance widths. If the resonance sweeps radially through phase space, stars may be captured or released independently of subhalo perturbations. Our treatment therefore neglects potentially important time-dependent effects that could either enhance or counteract subhalo-driven diffusion.

Despite these limitations, the broad conclusion of this work remains robust: the survival of bar-induced resonant features places meaningful constraints on the cumulative perturbative background in the local Galaxy. The novel framework presented here provides a foundation for more realistic modelling that relaxes these assumptions and includes higher order resonances.

\subsection{Future Work}

Promising extensions fall into three general categories:

First, the analytic framework can be expanded in several ways. A natural next step is to incorporate more complex time-dependent bar evolution, including secular bar slow-down, to capture more realistic Galactic evolution. The formalism can also be extended to higher-order resonances beyond co-rotation, which may act as more sensitive probes of low-mass subhalo perturbations. In addition, relaxing the assumption of purely impulsive encounters (by including adiabatic corrections and terms describing resonant coupling between subhaloes and the bar) will enable a more complete treatment of subhalo interactions. Finally, extending the present analysis from planar orbits to full three-dimensional phase space, including eccentric halo orbits and vertical structure, will allow us to probe more interesting and relevant scenarios.

Second, more advanced numerical simulations will play a key role in testing and refining this framework. In particular, tools such as \textit{StreamSculptor} \citep[][]{nibauer2025streamsculptor} offer a path toward modelling the response of a test-particle to a realistic population of subhaloes instead of a single subhalo. Beyond this, moving away from idealised test-particle experiments toward cosmological simulations will make it possible to apply and evaluate the method in environments that capture the full complexity of galaxy formation and evolution.

Third, observational studies will be essential to test these predictions. A clear target is to search for signatures of higher-order resonances (such as ultraharmonics and beyond) in observational data. Additionally, identifying an over-population or under-population at specific resonances relative to smooth dynamical expectations may provide the first empirical evidence of subhalo-driven diffusion in the Milky Way.

It is worth nothing that this framework could, in principle, be applied to external galaxies. While the challenges to such an endeavour are self-evident, the increasing precision with which the astronomy community measures extragalactic dynamical features \citep[e.g.][]{sola2025strrings} offers a glimpse into its practical utility.

\section*{Acknowledgements}

EYD thanks Shaunak Modak, Nathaniel Starkmann and Denis Erkal for helpful conversation, as well as the Cosmic Codebreakers group at MKI and the Streams group at Cambridge for useful feedback.

\section*{Data Availability}

The work in this project can be reproduced with publicly available software.



\bibliographystyle{mnras}
\bibliography{resonance_subhalo} 




\appendix

\section{Hernquist subhalo potential}

In this appendix we show the result of replacing the Plummer potential with the Hernquist potential \citep[][]{hernquist1990analytical} in the calculation of the diffusion coefficient $D_{\rm II}$. The form of the potential for a Plummer sphere is
\begin{equation}
    \Phi_{\rm H}(r) = -\frac{GM}{r + r_{\rm s}},
\end{equation}

where the scale radius can be related to the mass by \citep[][]{erkal2016number}
\begin{equation}
    r_{\rm s}(M) = 1.05 {\,\rm kpc} \left(\frac{M}{10^8 {\rm M}_{\odot}}\right)^{1/2}.
\end{equation}
The result of making this change is just a change to the value of $\kappa$ to $\kappa = 1.05^2 \times 10^{-8}$, which ultimately gives a value of $\mathcal{B}_0 = 28.02$.

\begin{figure}
    \centering
    \includegraphics[width=0.95\columnwidth]{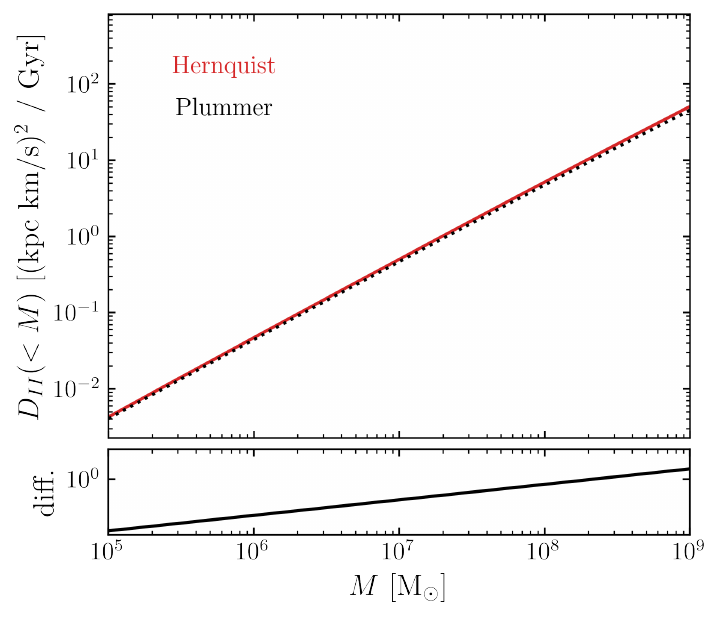}
    \caption{A comparison of the diffusion coefficient between a Hernquist potential and a Plummer potential. The top panel shows the diffusion coefficient itself, and the bottom panel shows the difference between the two cases. It is evident that the difference is minimal.}
    \label{fig:subhalo_type_compare}
\end{figure}

\section{Subhalo velocity distribution in the star's rest frame}\label{appendix:velocity}

In Sec.~\ref{sec:many_subhaloes} we require the distribution of relative speeds between the resonant star and an isotropic population of subhaloes. Here we outline the steps that lead from the isotropic Maxwellian distribution in the Galactic rest frame to the speed distribution $f(u)$ used in Eq.~\ref{eq:f(u)}.

We assume that subhalo velocities in the inertial frame follow a three-dimensional Maxwellian,
\begin{equation}
    f_{\bm{w}}(\bm{w}) = \left(\frac{1}{2\pi\sigma^{2}}\right)^{3/2} \exp\!\left[-\,\frac{|\bm{w}|^{2}}{2\sigma^{2}}\right],
\end{equation}
with one-dimensional dispersion $\sigma$.Because the transformation from $\bm{w}$ to $\bm{u}$ is a simple translation, the distribution of $\bm{u}$ is
\begin{equation}
f_{\bm{u}}(\bm{u})
  = \left(\frac{1}{2\pi\sigma^{2}}\right)^{3/2}
    \exp\!\left[-\,\frac{|\bm{u} + \bm{v}|^{2}}{2\sigma^{2}}\right].
\end{equation}
To obtain the speed distribution $f(u)$, we integrate over all directions of $\bm{u}$ at fixed magnitude $u$. Writing the relative angle between $\bm{u}$ and $\bm{v}$ as $\theta$,
\begin{equation}
|\bm{u} + \bm{v}|^{2} = u^{2} + v_\ast^{2} + 2 u v_\ast \cos\theta,
\end{equation}
and the angular integral becomes
\begin{equation}
f(u) = 2\pi u^{2}
       \left(\frac{1}{2\pi\sigma^{2}}\right)^{3/2}
       \int_{-1}^{+1}
       \exp\!\left[
         -\frac{u^{2} + v_\ast^{2} + 2u v_\ast \mu}{2\sigma^{2}}
       \right]
       d\mu ,
\end{equation}
where $\mu = \cos\theta$. Carrying out the integral yields the expression used in Sec.~\ref{sec:many_subhaloes}:
\begin{equation}
f(u)
 = \frac{u}{\sigma v_*\sqrt{2\pi}}
   \left[
     \exp\!\left(-\frac{(u-v_*)^{2}}{2\sigma^{2}}\right)
     - \exp\!\left(-\frac{(u+v_*)^{2}}{2\sigma^{2}}\right)
   \right].
\end{equation}
This function correctly normalises to unity and reduces to the isotropic Maxwellian limit when $v_*\rightarrow0$.

\section{Subhalo impact parameter distribution in star's rest frame}\label{appendix:impact}

Because the Fisher distribution is infrequently used in astronomy, we clarify details here. We require the expectation value of the squared azimuthal component of the impact parameter, $b_{\varphi}$, conditioned on the relative velocity direction $\hat{\bm{u}}$. Because the star moves with velocity $\bm{v}$, the relative-velocity distribution is no longer isotropic in 3D; instead, isotropy is preserved only in the two-dimensional plane perpendicular to $\bm{u}$. Here we show how this leads to Eq.~\ref{eq:b_correction}.

Let $\bm{b}$ be the three-dimensional impact parameter vector at closest approach, and let $\bm{e}_{\varphi}$ be the azimuthal unit vector of the star at encounter. The azimuthal component of the impact parameter is
\begin{equation}
b_{\varphi} = \bm{b} \cdot \bm{e}_{\varphi}.
\end{equation}
We decompose $\bm{e}_{\varphi}$ into components parallel and perpendicular to the relative velocity direction $\hat{\bm{u}}$:
\begin{equation}
\bm{e}_{\varphi}
   = (\bm{e}_{\varphi}\cdot\hat{\bm{u}})\,\hat{\bm{u}}
     + \bm{e}_{\varphi,\perp},
\qquad
\bm{e}_{\varphi,\perp}\cdot \hat{\bm{u}} = 0.
\end{equation}
Because $\bm{b}$ lies uniformly in the plane perpendicular to $\hat{\bm{u}}$ in the star’s rest frame, only the perpendicular component contributes on average:
\begin{equation}
\langle b_{\varphi}^{2} \mid \bm{u} \rangle
  = \frac{b^{2}}{2}\,|\bm{e}_{\varphi,\perp}|^{2}.
\end{equation}
Using $|\bm{e}_{\varphi}|^{2}=1$ and the above decomposition,
\begin{equation}
|\bm{e}_{\varphi,\perp}|^{2}
  = 1 - (\bm{e}_{\varphi}\cdot \hat{\bm{u}})^{2}.
\end{equation}
Because shifting by $\bm{v}$ breaks isotropy, the relative-velocity directions are biased toward the direction of $+\bm{v}$. The correct directional distribution is the Fisher distribution \citep[][]{fisher1953dispersion}, with concentration parameter:
\begin{equation}
\xi \equiv \frac{u v_*}{\sigma^{2}}.
\end{equation}
The probability density for a direction $\hat{\bm{u}}$ relative to the preferred direction $\bm{e}_{\varphi}$ is
\begin{equation}
f_{\hat{\bm{u}}}(\hat{\bm{u}})
  = \frac{\xi}{4\pi\sinh\xi}
    \exp\!\left(\xi\, \bm{e}_{\varphi}\cdot\hat{\bm{u}}\right).
\end{equation}
We must now evaluate
\begin{equation}
\big\langle (\bm{e}_{\varphi}\cdot\hat{\bm{u}})^{2}\big\rangle
   = \int (\bm{e}_{\varphi}\cdot\hat{\bm{u}})^{2}
       f_{\hat{\bm{u}}}(\hat{\bm{u}})\,d\Omega.
\end{equation}
The Fisher distribution has the identity
\begin{equation}
\langle \cos \theta \rangle
   = \frac{1}{\xi}\coth\xi - \frac{1}{\xi^{2}},
\end{equation}
where $\theta$ is the angle between $\hat{\bm{u}}$ and the preferred direction $\bm{e}_{\varphi}$. Thus
\begin{equation}
\big\langle (\bm{e}_{\varphi}\cdot\hat{\bm{u}})^{2} \big\rangle
 = \frac{1}{\xi}\coth\xi - \frac{1}{\xi^{2}}.
\end{equation}
Substituting into $1 - \langle(\bm{e}_{\varphi}\cdot\hat{\bm{u}})^{2}\rangle$ and multiplying by $b^{2}/2$ gives
\begin{equation}
\langle b_{\varphi}^{2} \mid \bm{u} \rangle
   = b^{2}
     \left[
       \frac{\coth\xi}{\xi}
       - \frac{1}{\xi^{2}}
     \right],
\end{equation}
which is Eq.~\ref{eq:b_correction} of the main text. In the limit $v_*\!\rightarrow 0$ (hence $\xi\rightarrow 0$), the directional bias vanishes, and the expression smoothly reduces to the isotropic value
\begin{equation}
\langle b_{\varphi}^{2} \rangle \to \frac{b^{2}}{3}.
\end{equation}

\section{Simulation potential}

\begin{figure}
    \centering
    \includegraphics[width=0.95\columnwidth]{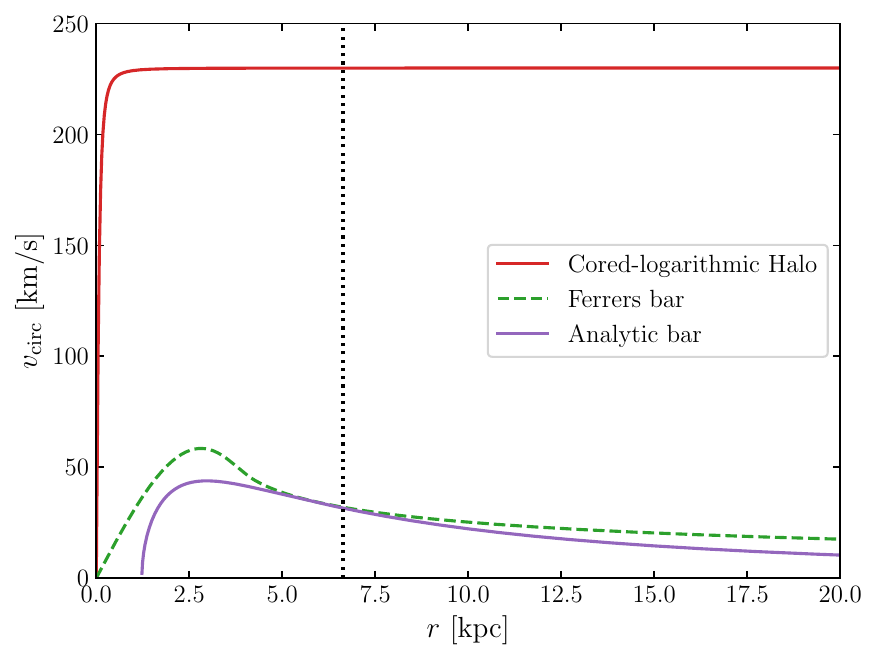}
    \caption{Rotation curves (along the major axis) for the cored-logarithmic halo potential, Ferrer's bar potential and analytic bar potential described in Sec.~\ref{sec:mathe_background}. The vertical dotted line shows the position of the co-rotation radius.}
    \label{fig:potential_vcirc}
\end{figure}

For clarity, in Fig.~\ref{fig:potential_vcirc} we compare the rotation curves (computed along the major axis) of the three potential components used throughout this work: the cored-logarithmic halo, the Ferrers bar, and the analytic bar model. Although the analytic work and the simulations rely on slightly different bar parametrisations, the figure illustrates that their contributions to the circular velocity along the bar’s major axis remain broadly consistent near the co-rotation radius. Therefore, the simulated resonant stars experience restoring forces and resonant conditions closely matching those assumed in the analytic framework, allowing for a like-for-like comparison between the impulse approximation and the test-particle experiments.

\bsp	
\label{lastpage}
\end{document}